\pdfoutput=1
\documentclass[iop]{emulateapj}

\slugcomment{}

\shorttitle{Study of star formation activity in S235}
\shortauthors{L.~K. Dewangan et al.}

\begin{document}

\title{A multi-wavelength study of star formation activity in the S235 complex}
\author{L.~K. Dewangan\altaffilmark{1,3}, D.~K. Ojha\altaffilmark{2}, A. Luna\altaffilmark{1}, B.~G. Anandarao\altaffilmark{3}, J.~P. Ninan\altaffilmark{2}, K.~K. Mallick\altaffilmark{2}, and Y.~D. Mayya\altaffilmark{1}}

\email{lokeshd@prl.res.in}

\altaffiltext{1}{Instituto Nacional de Astrof\'{\i}ísica, \'{O}ptica y Electr\'{o}nica, Luis Enrique Erro \# 1, Tonantzintla, Puebla, M\'{e}xico C.P. 72840}
\altaffiltext{2}{Department of Astronomy and Astrophysics, Tata Institute of Fundamental Research, Homi Bhabha Road, Mumbai 400 005, India}
\altaffiltext{3}{Physical Research Laboratory, Navrangpura, Ahmedabad - 380 009, India}

\begin{abstract}
We have carried out an extensive multi-wavelength study to investigate the star formation process 
in the S235 complex. The S235 complex has a sphere-like shell appearance at wavelengths 
longer than 2 $\mu$m and harbors an O9.5V type star approximately at its center.
Near-infrared extinction map traces eight subregions (having A$_{V}$ $>$ 8 mag), and 
five of them appear to be distributed in an almost regularly spaced manner along the sphere-like shell surrounding the ionized emission. 
This picture is also supported by the integrated $^{12}$CO and $^{13}$CO intensity maps and by 
Bolocam 1.1 mm continuum emission. The position-velocity analysis of CO reveals 
an almost semi-ring like structure, suggesting an expanding H\,{\sc ii} region. 
We find that the Bolocam clump masses increase as we move away from the location of the ionizing star.
This correlation is seen only for those clumps which are distributed near the edges of the shell. 
Photometric analysis reveals 435 young stellar objects (YSOs), 59\% of which are found in clusters. 
Six subregions (including five located near the edges of the shell) are very well correlated with the dust clumps, CO gas, and YSOs.
The average values of Mach numbers derived using NH$_{3}$ data 
for three (East~1, East~2, and Central~E) out of these six subregions are 2.9, 2.3, and 2.9, indicating these subregions are supersonic. 
The molecular outflows are detected in these three subregions, further confirming the on-going star formation activity.
Together, all these results are interpreted as observational evidence of positive feedback of a massive star. 
\end{abstract}
\section{Introduction}
\label{sec:intro} 
The energetics of massive stars can strongly influence the surroundings \citep{zinnecker07}. 
Massive stars can destroy star-forming clouds (i.e., negative feedback) and can also 
trigger star formation (i.e., positive feedback), leading to the formation of a new generation 
of stars including young massive star(s) \citep{deharveng10}. 
However, the feedback processes of massive stars are still poorly understood.

The extended star-forming region S235 is known to be a 
part of the giant molecular cloud G174+2.5 in the Perseus Spiral Arm \citep[e.g.][]{heyer96} and contains two known sites ``S235 complex" and ``S235AB region" 
(see Figure~1 in \citet{dewangan11} and also in \citet{kirsanova14}). The present work is focused on ``S235 complex" and does not include ``S235AB region". Different values of the distance (1.36 kpc, 1.59 kpc, 1.8 kpc, 2.1 kpc, and 2.5 kpc) to the extended star-forming 
region S235 are reported in the literature \citep[e.g.][]{georgelin73,israel78,evans81a,brand93,burns15,foster15}. 
In the present work, we have chosen a distance of 1.8 kpc following \citet{evans81a}, which is an intermediate value of the published distance range. 
The H\,{\sc ii} region associated with the S235 complex is predominantly ionized by a single massive star BD+35$^{o}$1201 
of O9.5V type \citep{georgelin73}. 
The S235 complex has been studied using multiple datasets spanning near-infrared (NIR) to radio wavelengths. 
The S235 complex is known as an active site of star formation, harboring young stellar clusters 
\citep[e.g.,][]{kirsanova08,dewangan11,camargo11,chavarria14} associated with known star-forming subregions, 
namely East~1, East~2, and Central \citep[e.g.][]{kirsanova08}.  In our previous work on the S235 complex 
using the {\it Spitzer}-IRAC data \citep[][hereafter Paper I]{dewangan11},  
we detected several young stellar objects (YSOs) including a High Mass Protostellar Object (HMPO) candidate 
as well as signatures of outflow activities. Using $^{13}$CO (1$-$0) line data, \citet{kirsanova08} found three molecular gas components 
(i.e., $-18$ km s$^{-1}$ $< V_{lsr} < -15$ km s$^{-1}$ (red), $-21$ km s$^{-1}$ $< V_{lsr} < -18$ km s$^{-1}$ (central), 
and $-25$ km s$^{-1}$ $< V_{lsr} < -21$ km s$^{-1}$ (blue)) in the direction of the S235 complex. 
However, the complex is well traced in mainly two molecular gas components (central and blue). 
More recently, \citet{kirsanova14} derived physical parameters of dense gas (i.e., gas density and temperature) 
in subregions of the complex using ammonia (NH$_{3}$) line observations. 
However, the properties of dense gas are not explored with respect to the ionizing star location. 
Previous studies indicated that the S235 H\,{\sc ii} region is interacting with its surrounding molecular cloud 
and the S235 complex has been cited as a possible site of triggered star formation \citep{kirsanova08,kirsanova14,camargo11}.  

{\it Spitzer} images revealed that the complex has a sphere-like shell morphology and 
the ionizing star is approximately located at its center (see Figure~\ref{fig1a} given in Paper~I). 
In addition to the observed interesting morphology, the complex is a relatively nearby star-forming site, making it a promising site to 
study the feedback of a massive star. 
In spite of the numerous existing observations and interpretations, the feedback of an O9.5V type star 
is not systematically explored in the S235 complex. 
The aim of the present work is to study the physical processes governing the interaction and feedback effect of a 
massive star on its surroundings.

In order to address above aim, we revisited the S235 complex using high sensitivity United Kingdom Infra-Red Telescope (UKIRT) 
Infrared Deep Sky Survey (UKIDSS) NIR data, Giant Metre-wave Radio Telescope (GMRT) 610 MHz radio continuum map, and 
dust continuum 1.1 mm data, in conjunction with published narrow band H$_{2}$ map, 
{\it Spitzer} mid-infrared (MIR) data, NH$_{3}$ line data, NRAO VLA Sky Survey (NVSS) 1.4 GHz continuum map, and CO line data.
We performed a detailed study of the distribution and kinematics of molecular gas in the complex. 
In order to systematically explore the feedback of a massive star, we estimated various pressure components 
(such as pressure of an H\,{\sc ii} region, radiation pressure, stellar wind pressure, pressure exerted 
by the self-gravitating molecular cloud, and ratio of thermal to non-thermal gas pressure).  
For a detailed study of the embedded young population in the complex, we employed different color-color and color-magnitude 
diagrams obtained using NIR and MIR data, as well as extinction map generated from NIR data, and the surface density analysis. 
Additionally, the physical properties of gas derived using the line and continuum data were investigated. 
 
In Section~\ref{sec:obser}, we provide the description of various datasets 
along with reduction procedures. In Section~\ref{sec:data}, we summarize the results related to the 
physical environment and point-like sources. 
The possible star formation scenario is discussed in Section~\ref{sec:disc}. 
Our main conclusions are summarized in Section~\ref{sec:conc}.
\section{Data and analysis}
\label{sec:obser}
Multi-wavelength data are employed to study the physical conditions in the S235 complex. 
The size of the selected region is $\sim$17$\farcm$6  $\times$ 15$\farcm$5, 
centered at $\alpha_{2000}$ = 05$^{h}$40$^{m}$57.8$^{s}$,
$\delta_{2000}$ = +35$\degr$51$\arcmin$13$\arcsec$, corresponding to a physical scale of about  9.2 pc  $\times$ 8.1 pc 
at a distance of 1.8 kpc.
\subsection{Narrow-band H$_{2}$ image}
Narrow-band H$_{2}$ ($\nu$ = $1-0$ S(1) at $\lambda$ = 2.122 $\mu$m ($\Delta \lambda =0.032\, \mu \rm m$)) imaging 
data of the S235 complex were obtained from 
the survey of extended H$_{2}$ emission \citep{navarete15}, conducted at the Canada-France-Hawaii Telescope (CFHT), 
Mauna Kea, Hawaii, using Wide-field InfraRed Camera (WIRCam). The observations were performed with the average seeing of $\sim$0$\farcs$7. 
The survey also provided K-band continuum ($\lambda$ = 2.218 $\mu$m; $\Delta \lambda =0.033\, \mu \rm m$) images, 
which were used to get the final continuum-subtracted H$_{2}$ map. 
Note that the H$_{2}$ image published by \citet{navarete15} contains only the East~2 and Central star-forming subregions \citep[see source IDs 164 and 165 in Table B2 given in][]{navarete15}, 
while the H$_{2}$ image in this work is presented for a larger area $\sim$10$\farcm$7 $\times$ 9$\farcm$9 toward the complex. 
One can find more details about the survey in the work of \citet{navarete15}. 
\subsection{Near-infrared data}
We obtained deep NIR photometric {\it JHK} images and the magnitudes of point sources in the region ($\sim$17$\farcm$6  $\times$ 15$\farcm$5) from 
the UKIDSS 6$^{th}$ archival data release (UKIDSSDR6plus) of the Galactic Plane 
Survey \citep[GPS;][]{lawrence07}. The survey uses the UKIRT Wide Field Camera 
\citep[WFCAM;][]{casali07}. 
Following the selection conditions given in \citet{lucas08},  
we retrieved only reliable NIR sources in the region. 
One can also find more details about the selection procedure of the GPS photometry in the work of \citet{dewangan15b}.
Two Micron All Sky Survey \citep[2MASS;][]{skrutskie06} data were obtained for bright sources that were saturated in the GPS catalog. 
We found 2444 sources detected in all the three NIR ({\it JHK}) bands. 
Additionally, 326 sources were selected having detections only in the H and K bands.
\subsection{{\it Spitzer} data}
The photometric images (3.6--24 $\mu$m) of the S235 complex were obtained from the {\it Spitzer} Space Telescope 
Infrared Array Camera \citep[IRAC;][]{fazio04} and the 
Multiband Imaging Photometer \citep[MIPS;][]{rieke04}. 
The final processed IRAC images (3.6--8.0 $\mu$m) and photometry of point sources were taken from Paper I. 
MIPS 24 $\mu$m observations were retrieved from the {\it Spitzer} public 
archive\footnote[1]{see http://sha.ipac.caltech.edu/applications/Spitzer/SHA/}, 
which were carried out in MIPS scan mode on 15 April 2008 (Program id 40005; PI: Giovanni Fazio).
MIPS mosaic at 24 $\mu$m was generated using the Basic Calibrated Data (BCD) images.
\subsection{Dust continuum 1.1 mm data}
Bolocam 1.1 mm image \citep{aguirre11} and Bolocam source catalog at 1.1 mm \citep[v2.1;][]{ginsburg13} were 
obtained from Bolocam Galactic Plane Survey (BGPS). 
The effective full width at half maximum (FWHM) of the 1.1 mm map is $\sim$33$\arcsec$. 
\subsection{Radio continuum data}
We used the archival radio continuum data at 610 MHz (50 cm) and 1.4 GHz (21 cm).
The 1.4 GHz map was retrieved from the NVSS archive. 
The beam size of the NVSS image is $\sim$45$\arcsec$ \citep{condon98}.
The  610 MHz continuum data were observed on 18--19 June 2005 (Project Code: 08SKG01) 
and were retrieved from the GMRT archive. 
GMRT radio data reduction was carried out using AIPS software, in a manner similar to that described in \citet{mallick13}. 
Since the S235 complex was away from the center of the observed field, 
the primary beam correction for 610 MHz was also carried out, using the AIPS PBCOR task and parameters 
from the GMRT manual\footnote[2]{http://gmrt.ncra.tifr.res.in/gmrt\_hpage/Users/doc/obs\_manual.pdf}. 
The synthesized beam size of the final 610 MHz map is $\sim$48$\arcsec$ $\times$ 44\farcs2. 
\subsection{Molecular CO line data}
The J=1$-$0 lines of $^{12}$CO and $^{13}$CO data were observed from the Five College Radio Astronomy 
Observatory (FCRAO) 14 meter telescope in New Salem, Massachusetts. 
The FCRAO beam sizes were 45$\arcsec$ and 46$\arcsec$ for $^{12}$CO and $^{13}$CO, respectively. 
The S235 complex was observed as part of the Extended Outer Galaxy Survey \citep[E-OGS,][]{brunt04}, 
that extends the coverage of the FCRAO Outer Galaxy Survey \citep[OGS,][]{heyer98} to Galactic longitude $l$ = 193$\degr$, 
over a latitude range of $-$3$\degr$.5 $\leq$ $b$ $\leq$ +5$\degr$.5. 
However, the data cubes of S235 complex were further re-processed and a document describing the re-processing 
data methods is given in \citet{brunt04} (also Brunt C.M. et al.; in preparation).
These CO data cubes were obtained from M. Heyer and C. Brunt (through private communication). 
\subsection{Other Data}
We utilized the publicly available archival WISE\footnote[3]{Wide Field Infrared Survey Explorer, which is a joint project of the
University of California and the JPL, Caltech, funded by the NASA} \citep{wright10} image at 12 $\mu$m (spatial resolution $\sim6\arcsec$). 
We also used previously published NH$_{3}$ line data from \citet{kirsanova14}. 
\section{Results}
\label{sec:data}
\subsection{Mid-infrared and radio continuum images}
\label{subsec:mir}
\begin{figure*}
\epsscale{0.62}
\plotone{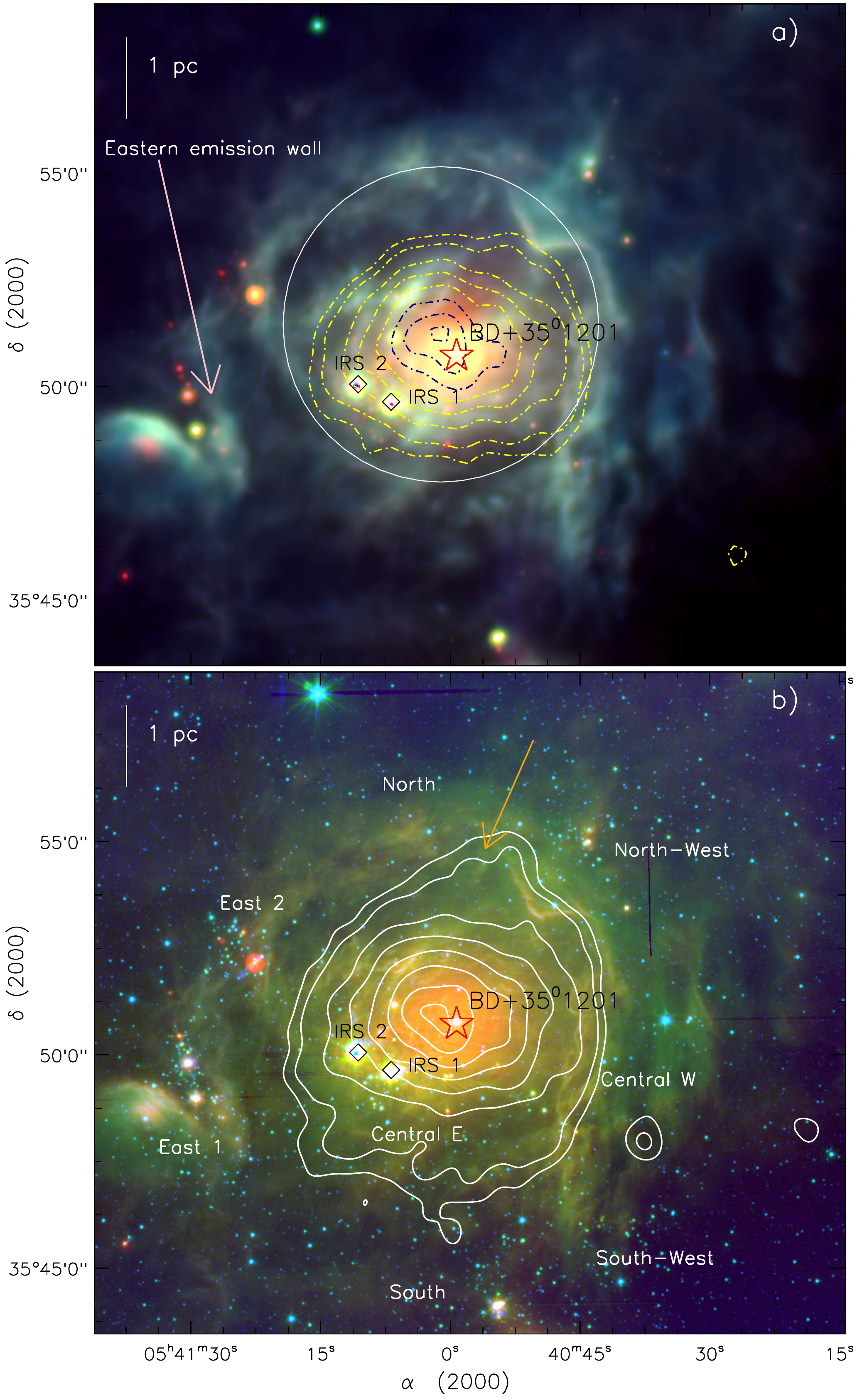}
\caption{\scriptsize MIR and NIR emissions toward the S235 complex (size of the selected 
field $\sim 17$\farcm$6 \times 15$\farcm$5$ (or 9.2 pc $\times$ 8.1 pc); 
centered at $\alpha_{2000}$ = 05$^{h}$ 40$^{m}$ 57.8$^{s}$, 
$\delta_{2000}$ = +35$\degr$ 51$\arcmin$ 13$\arcsec$). 
a) The image is the result of the combination of three bands: 24 $\mu$m in red (MIPS), 12.0 $\mu$m in 
green (WISE), and 8.0 $\mu$m in blue (IRAC).
Contours of NVSS 1.4 GHz radio continuum emission (beam size $\sim$45$\arcsec$) are superimposed with 5, 10, 25, 40, 55, 70, 85, and 98\% of 
the peak value (i.e., 0.124 Jy/beam).  The big circle highlights a sphere-like shell morphology seen in {\it Spitzer} images. 
A nebular emission is designated as eastern emission wall (see Section~\ref{subsec:mir} for more details). 
b) Color-composite map using {\it Spitzer} MIPS 24 $\mu$m (red), IRAC 4.5 $\mu$m (green), and 
UKIDSS K (blue) images. 
GMRT 610 MHz radio continuum contours (beam size $\sim$48$\arcsec$ $\times$ 44\farcs2) are drawn 
in white color with levels of 5, 10, 25, 40, 55, 70, 85, and 98\% of the peak value (i.e., 0.158 Jy/beam). 
Different subregions are also labeled in the map (see text for more details). 
In both the panels, the positions of two bright sources (IRS 1 and IRS 2) and an O9.5V star (BD+35$^{\degr}$1201) are marked with diamond and 
star symbols, respectively. 
The scale bar on the top left shows a size of 1 pc at a distance of 1.8 kpc in both the figures.}
\label{fig1a}
\end{figure*}

The most prominent feature of the S235 complex observable in the infrared regime, the sphere-like 
shell morphology, is detected at wavelengths longer than 2 $\mu$m (see Figure~\ref{fig1a}). 
Figure~\ref{fig1a}a is a three-color composite image made using MIPS 24 $\mu$m in red, WISE 12.0 $\mu$m in green, 
and IRAC 8.0 $\mu$m in blue. Figure~\ref{fig1a}b shows a three-color composite image using 
MIPS 24 $\mu$m (red), IRAC 4.5 $\mu$m (green), and UKIDSS 2.2 $\mu$m (blue). 
Radio continuum contours at 1.4 GHz and 610 MHz are also overlaid on the color-composite images in Figures~\ref{fig1a}a and~\ref{fig1a}b, respectively.
These radio continuum contours trace the ionized emission in the S235 complex. 
The distribution of ionized emission is almost spherical in both the radio maps.
The location of a previously characterized O9.5V star (BD +35$^{o}$1201) appears at the peak of radio continuum emission. 
The bulk of the ionized emission, as traced at the sensitivity levels shown in the Figures~\ref{fig1a}a and~\ref{fig1a}b, is well located inside 
the sphere-like shell, as highlighted by a circle in Figure~\ref{fig1a}a. 
The radio emission at 610 MHz appears elongated in north direction, as highlighted by an 
arrow (see Figure~\ref{fig1a}b). This particular feature does not appear to be seen in the  NVSS 1.4 GHz map. 
The WISE 12 $\mu$m band is dominated by the 11.3 $\mu$m polycyclic aromatic hydrocarbon (PAH) emission  
as well as the warm dust continuum emission. The 24 $\mu$m image shows the warm dust emission,
whereas the 8 $\mu$m band contains PAH emission features at 7.7 and 8.6 $\mu$m (including the continuum). 
In Figure~\ref{fig1a}, we notice that the warm dust emission is surrounded by the 8 $\mu$m emission. 
Furthermore, the correlation of the warm dust and ionized emission is evident, which has generally been found in H\,{\sc ii} 
regions \citep[e.g.][]{deharveng10,paladini12}.  
In addition to the extended emission, the embedded stellar contents are visually seen around the sphere-like shell. 
The different subregions (viz. East~1, East~2, North, North-West, Central~W, Central~E, South-West, and South) in the S235 complex are
shown in Figure~\ref{fig1a}b. Note that three subregions (East~1, South-West, and South) are located away from the edges of the 
sphere-like shell (see Figure~\ref{fig1a}). 
There is noticeable nebular emission seen between East~1 subregion and the sphere-like shell, 
which is designated as eastern emission wall in Figure~\ref{fig1a}a.
The Central~E subregion contains two bright sources, namely EB IRS~1/G173.6328+02.8064 (hereafter, IRS~1) 
and EB IRS~2/G173.6339+02.8218 (hereafter, IRS~2) \citep{evans81a} with previously estimated 
luminosities of $\sim$2.5 $\times$ 10$^{3}$ L$_{\odot}$ 
and $\sim$1.5 $\times$ 10$^{3}$ L$_{\odot}$ \citep{evans81b}, respectively. 
Furthermore, radio continuum emissions are absent toward these two sources (IRS~1 and IRS~2), as pointed out 
by \citet{nordh84}. Combining these information with the radio continuum data presented in this work, 
one can note that no radio continuum emission peaks are seen toward the subregions. 
Figure~\ref{fig1a}b shows the enhanced 4.5 $\mu$m and K-band extended emission features in the East~1 and East~2 
subregions. In Paper~I, the signatures of outflow activities were found toward East~1 and East~2 subregions 
due to the presence of shock-excited H$_{2}$ emissions as revealed by the IRAC ratio maps. 
Note that the embedded sources in East~1 subregion appear to be separated by the eastern emission 
wall from the sphere-like shell (see Figure~\ref{fig1a}a).
\subsection{Near-infrared extinction map}
\begin{figure*}
\epsscale{0.68}
\plotone{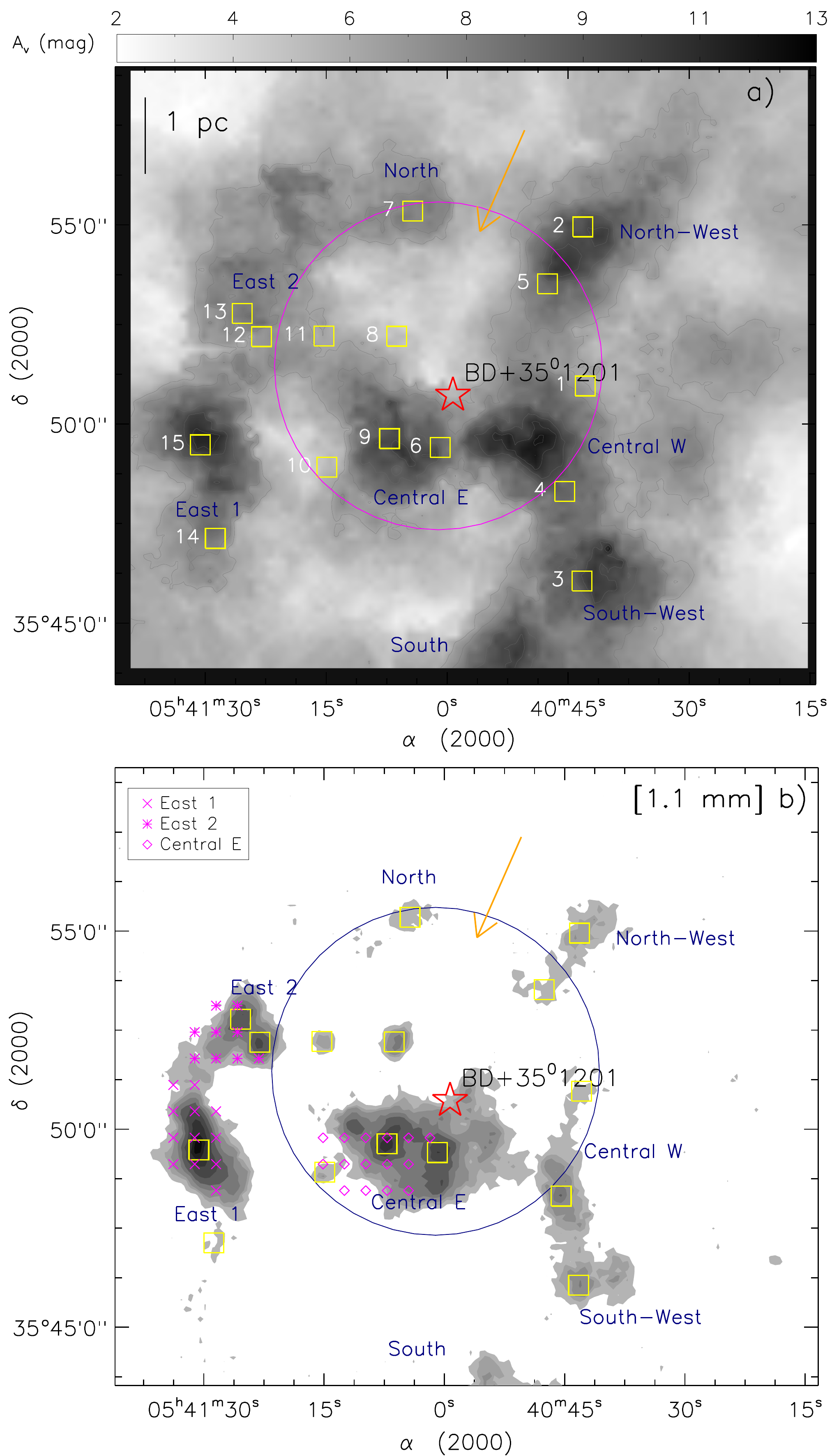}
\caption{\scriptsize a) Visual extinction map of the S235 complex obtained using the NIR data. 
b) Contour map of Bolocam 1.1 mm dust continuum emission (beam size $\sim$33$\arcsec$).
The contour levels are 8, 15, 20, 30, 40, 55, 70, 85, and 95\% of the peak value i.e. 1.2 Jy/beam.
The positions of previously observed NH$_{3}$ line observations in East~1 (x), East~2 (*), 
and Central~E ($\diamond$) subregions are marked (see text for details and also Table~\ref{tab3}). 
The positions of dust clumps at 1.1 mm are marked by yellow square symbols in both the 
panels and are labeled in panel ``a" (see Table~\ref{tab1}).
In both the panels, other marked symbols and labels are similar to those shown in Figure~\ref{fig1a}.}
\label{fig2a}
\end{figure*}

The visual extinction (A$_{V}$) map of the S235 complex is generated using the publicly available GPS NIR (JHK) photometric data. 
The extinction value of individual stars was computed using a color-excess 
along the line of sight \citep[$E(H - K) = (H - K)_{obs} - <(H - K)_{0}>$; see equation given in][]{lada94} using the 
reddening laws of \citet{indebetouw05}. The $<$(H - K)$_{0}$$>$ is the mean intrinsic color of the stars, 
which was determined to be $\sim$0.25 from the nearby control 
field (similar size as target region; central coordinates: $\alpha_{J2000}$ = 05$^{h}$43$^{m}$35$^{s}$.8, 
$\delta_{J2000}$ = +35$\degr$20$\arcmin$20$\arcsec$.4). 
The extinction map was created by mean value of 40 nearest neighbours. 
The resultant A$_{V}$ map of S235 complex is shown in Figure~\ref{fig2a}a, 
which allows to trace several embedded subregions in the complex. 
All these subregions are indicated in Figure~\ref{fig2a}a, which are already 
highlighted in Figure~\ref{fig1a}b based on the visual appearance of stellar contents. 
The A$_{V}$ values vary between 2 and 12.5 mag with an average of about 5.8 mag. 
The A$_{V}$ map also resembles the sphere-like morphology as highlighted in Figure~\ref{fig1a}a, 
except in the north direction, where it seems to be broken (shown by an arrow in Figure~\ref{fig2a}a). 

Fifteen Bolocam dust continuum clumps at 1.1 mm were retrieved from Bolocam source catalog (v2.1) 
and are shown in Figure~\ref{fig2a}a. Figure~\ref{fig2a}a illustrates a correlation between extinction and dust clumps. 
Most of the clumps are located in the region having A$_{V}$ greater than 8 mag. 
\subsection{Clump properties}
\label{subsec:dstmas}
Dust continuum emission is known to trace the dense and cold regions. 
In Figure~\ref{fig2a}b, we show Bolocam dust continuum emission at 1.1 mm, which is seen toward all the marked subregions in Figure~\ref{fig1a}b.
It is important to note that previously published SCUBA/JCMT 850 $\mu$m data (beam size $\sim$14$\arcsec$) 
are not available toward the ``North" and ``North-West" subregions of the S235 
complex \citep[e.g.][]{klein05,francesco08}. 
Therefore, the 1.1 mm data (beam size $\sim$33$\arcsec$) provide a more complete information 
of the dense and cold regions of the S235 complex. 
The identified fifteen clumps in the 1.1 mm map are also presented in Figure~\ref{fig2a}b. 
The association of clump(s) with each subregion is listed in Table~\ref{tab1}. Five subregions 
(i.e., North-West, Central~W, Central~E, East~1, and East~2) contain at least two clumps, while other 
two subregions (i.e., South-West and North) are associated with a single clump. 
In order to derive gas mass M$_{g}$ for each clump at 1.1 mm emission, 
the following formula has been utilized \citep{hildebrand83}:
\begin{equation}
M_{g} \, = \, \frac{D^2 \, S_\nu \, R_t}{B_\nu(T_d) \, \kappa_\nu}
\end{equation} 
\noindent where $S_\nu$ is the integrated 1.1 mm flux (Jy), 
$D$ is the distance (kpc), $R_t$ is the gas-to-dust mass ratio (assumed to be 100), 
$B_\nu$ is the Planck function for a dust temperature $T_d$, 
and $\kappa_\nu$ is the dust absorption coefficient.  
In the calculation, we use $\kappa_\nu$ = 1.14\,cm$^2$\,g$^{-1}$ \citep[e.g.][]{enoch08,bally10},  
$D$ = 1.8 kpc, and $T_d$ = 20 K 
\citep[an average temperature value estimated from NH$_{3}$ line data; see][for more details]{kirsanova14}. 
The masses computed from the observed dust continuum data are listed in Table~\ref{tab1}. 
The total gas mass of the 15 clumps comes out to be $\sim$1179 M$_{\odot}$. 
The clump masses vary between 7 M$_{\odot}$ and 285 M$_{\odot}$.   
All the clumps have masses below 100 M$_{\odot}$ except three clumps.
These three clumps are associated with Central~E (clump IDs 6 and 9) and East~1 (clump ID 15) 
subregions and have masses more than 200 M$_{\odot}$ 
(see Table~\ref{tab1}). In an earlier work of the S235 complex (excluding North and North-West subregions), 
\citet{klein05} selected thirteen dust emission clumps in the JCMT SCUBA 850 $\mu$m survey 
\citep[beam size $\sim$14$\arcsec$; see Fig.~1 of IRAS 05377+3548 in][]{klein05}. 
They estimated a mass range of clumps between 10 and 120 M$_{\odot}$, using an average dust temperature of 
20 K, a gas-to-dust mass ratio of 150, and a distance of 1.8 kpc. 
\citet{kirsanova14} also computed a mass range of dense gas in the clumps between 12 and 
250 M$_{\odot}$ \citep[see Table~6 in][]{kirsanova14}. Considering the clump masses estimated in the present work, 
one can find a noticeable difference in the clump mass at the high end of the estimated range compared to 
the previous works. At the low end of the range, the gas mass is almost same in the present calculations compared to the earlier 
works. More extensive high sensitivity and high resolution observations at mm/sub-mm wavelengths would 
allow to further explore the clump mass at the high end of the estimated mass range.

In Figure~\ref{fig3a}, we show the distribution of clump masses as a function of distance from the location of the O9.5V star. 
Interestingly, we notice that the clump masses increase as we move away from the location of the ionizing star. 
This correlation is seen only for those clumps which are distributed near the edges of the sphere-like 
shell (e.g. Central~E, East~2, North, and North-West subregions).  
The implication of this correlation for the star formation process is discussed in Section~\ref{sec:disc}. 

The column densities of clumps can also be computed using the Bolocam data.
The column density per beam is given by \citep[see][]{deharveng10}: 
\begin{equation}
N(\mathrm H_2) =
\frac{S_\nu \, R_t}
{\kappa_\nu\,\,
B_\nu(T_d)\,\,
\mu_{H2}\,\,m_{\mathrm H}\,\,\Omega_\mathrm{beam}}\mathrm, 
\end{equation}
\noindent In the equation above, S$_\nu$, $R_t$, $\kappa_\nu$, and $B_\nu$(T${_d}$) are defined as in Equation 1, 
$\mu_{H2}$ = 2.8 is the mean molecular weight per hydrogen molecule \citep{kauffmann08}, 
the hydrogen atom mass $m_{\mathrm H}$ is in grams, and the beam solid angle 
$\Omega_\mathrm{beam}$ (= $\pi \, (FWHM / (206265 \times \sqrt{4\,ln\,2}))^{2}$ = $2.9 \times 10^{-8}$ 
for Bolocam beam FWHM= 33$\arcsec$) is in steradians, and 
$N(\mathrm H_2)$ \citep[= $2.0 \times 10^{22}\,\,S_\nu (Jy)$ at $T_D$ = 20 K; see][]{bally10} is 
per square centimeter. The column densities thus computed are listed in Table~\ref{tab1}. Bolocam clumps have column densities 
between 0.33 $\times$ 10$^{22}$ and 13.5 $\times$ 10$^{22}$ cm$^{-2}$.
\subsection{IRAC ratio and H$_{2}$ maps}
\label{subsec:h2out}
In order to trace molecular H$_{2}$ emission in the complex, we examined the {\it Spitzer}-IRAC ratio (4.5 $\mu$m/3.6 $\mu$m) 
map and the narrow H$_{2}$ (2.12 $\mu$m) map. The {\it Spitzer}-IRAC ratio maps have been utilized by many authors to 
study the interaction between a massive star and its surrounding interstellar
medium (ISM) \citep{povich07,watson08,kumarld10a,kumarld10,dewangan11,dewangan12}.
IRAC 3.6 $\mu$m and 4.5 $\mu$m bands have almost identical 
point response function (PRF), therefore, one can directly take the ratio of 4.5 $\mu$m to 3.6 $\mu$m bands.
In Figure~\ref{fig4ua}, we display a ratio map of 4.5 $\mu$m/3.6 $\mu$m emission. 
The spherical shell-like morphology is evident and the map depicts the edges of the shell. 
Additionally, several features are seen that are present toward the edges of the shell and in the vicinity of a massive star. 
In general,  the bright emission region in ratio 4.5 $\mu$m/3.6 $\mu$m map indicates the excess 4.5 $\mu$m emission, 
while the remaining black or dark gray regions suggest the domination of 3.6 $\mu$m emission.  
IRAC 4.5 $\mu$m band contains a prominent molecular hydrogen line emission ($\nu$ = 0--0 $S$(9); 4.693 $\mu$m), 
which can be excited by outflow shocks, and a 
hydrogen recombination line Br$\alpha$ (4.05 $\mu$m).
IRAC 3.6 $\mu$m band harbors PAH emission at 3.3 $\mu$m as well as a prominent molecular hydrogen feature at 3.234 
$\mu$m ($\nu$ = 1--0 $ O$(5)). In ratio map, we also notice a bright region in the vicinity of the massive star 
that coincides with warm dust emission traced at 24 $\mu$m as well as with the peak of the radio emission (see arrow in Figure~\ref{fig4ua}). 
This bright emission region probably traces the Br$\alpha$ feature originated by photoionized gas. 
In East~1, East~2, Central~E, and North-West subregions, we found bright emissions that are marked in the ratio map. 
Note that there are no radio continuum emission detected toward these subregions. Therefore, these emissions are 
probably H$_{2}$ features tracing outflow activities (see Section~\ref{subsec:vel} for molecular outflows). 
Figure~\ref{fig5} shows a continuum-subtracted 2.12 $\mu$m H$_{2}$ ($\nu$ = $1-0$ S(1)) image, 
revealing the presence of H$_{2}$ emission in the complex. 
The complex contains diffuse or filamentary-like H$_{2}$ features as well as extended H$_{2}$ 
emission. The extended H$_{2}$ emission is seen as polar 
structures (i.e., monopolar, bipolar, and multipolar; see East~1, East~2, and Central~E subregions in Figures~\ref{fig5} and~\ref{fig6}).
The bipolar emission structure is often interpreted as a bipolar outflow excited by a young star.
The multipolar structures could be originated due to presence of bipolar outflows associated with multiple young stars.
The monopolar H$_{2}$ structure could be explained by the presence of a bipolar outflow excited by a young star, 
however its red-shifted component is hidden from view due to inclination angle, extinction, and 
opacity of the ISM around the young star. 
In Figure~\ref{fig6}, we present H$_{2}$, 4.5 $\mu$m, and 24 $\mu$m images of three 
subregions (i.e., East~1,  East~2, and Central-E), where polar structures are traced in the H$_{2}$ map. 
In East~1 subregion, one multipolar structure is clearly traced. Additionally, some H$_{2}$ knots are seen. 
The H$_{2}$ map confirms intense outflow or jet activities in this subregion. 
The extended 4.5 $\mu$m emission is also detected similar to those seen in the H$_{2}$ map. 
The prominent 4.5 $\mu$m emission was reported in Paper~I using the IRAC ratio map 4.5 $\mu$m/8.0 $\mu$m 
(see Figure~11 in Paper~I). 
This subregion contains a cluster of embedded YSOs which makes difficult to pinpoint the exact exciting source(s) of outflows.
However, some probable driving sources (i.e. YSOs) of outflows, that are taken from Paper~I, are marked in the 4.5 $\mu$m and 24 $\mu$m images. 
These highlighted YSOs appear bright in the 24 $\mu$m image. 
In East~2 subregion, a bipolar structure is detected and is also seen in the 4.5 $\mu$m image (also see Figure~12 in Paper~I). 
In Paper~I, the driving source of this outflow was characterized as a HMPO candidate. 
The position of this YSO is also marked in the 4.5 $\mu$m and 24 $\mu$m images. 
The YSO is very bright in the 24 $\mu$m image. 
In Central-E subregion (see Figure~\ref{fig6}), we find a bipolar and two monopolar structures 
\citep[also see source IDs 164 and 165 in Table B2 given in][]{navarete15}. 
The probable exciting sources of one bipolar structure (Outflow-ce1) 
and one monopolar structure (Jet-ce) are marked in the 4.5 $\mu$m and 24 $\mu$m images (see Figure~\ref{fig6}). 
The 24 $\mu$m counterparts of these driving sources are also seen. 
However, the driving source of one monopolar structure (Outflow-ce2; Figure~\ref{fig6}) is not detected.  
In general, one can refer H$_{2}$ nebulosity as an outflow when it traces back to a source. 
In the present case, considering the morphology of H$_{2}$ nebulosity \citep[e.g.][]{takami10}, 
we refer this monopolar structure as the outflow signature. 
It is generally accepted that the molecular outflows directly indicate the presence of star formation processes. 
A comparison of the 2.12 $\mu$m H$_{2}$ emission with the features detected in the ratio map suggest that the black 
region in the ratio map probably traces the H$_{2}$ features (see Figures~\ref{fig4ua} and~\ref{fig5}). 
However, the presence of 3.3 $\mu$m PAH feature can not be ignored. 
The prominent diffuse or filamentary-like H$_{2}$ features traced in the H$_{2}$ map could be explained by ultra-violet (UV) fluorescence.  
In summary, the IRAC ratio and H$_{2}$ maps trace photodissociation regions (or photon-dominated regions, 
or PDRs) around the H\,{\sc ii} region and star formation activities in the complex.
\subsection{CO molecular gas associated with the S235 complex}
\label{subsec:vel}
In the following, we present molecular line data analysis, which enables us to examine the ongoing physical 
processes and velocity structures present in the complex.
\subsubsection{$^{12}$CO and $^{13}$CO Distributions}
In Figure~\ref{fig7}, we present the velocity channel maps of the J = 1$-$0 line of $^{13}$CO (at intervals of 0.25 km s$^{-1}$). 
The maps reveal the distribution of molecular gas components along the line of sight. 
The maps are shown for the velocity range of $-$22.75 km s$^{-1}$ to $-$15 km s$^{-1}$. 
One may notice from Figure~\ref{fig7} that the distribution of molecular gas is clumpy, as also previously pointed out by \citet{heyer96}. 
In general, the $^{13}$CO line is more optically thin compared to $^{12}$CO line. 
Therefore, the $^{13}$CO line data can trace dense condensation and its associated velocity better than $^{12}$CO. 
Considering this fact, we preferred to show only $^{13}$CO channel maps of the complex. 
In the velocity range from $-$22.75 to $-$21.25 km s$^{-1}$, Central~W, East~2, and North-West subregions are well detected.  
In the velocity range from $-$21 to $-$19.75 km s$^{-1}$, Central~E, Central~W, East~1, East~2, North, and North-West subregions are traced.  
In the velocity range from $-$19.50 to $-$18.0 km s$^{-1}$, Central~E and East~1 subregions are seen. 
The $^{13}$CO gas kinematics presented in this work is consistent with the previous work of \citet{kirsanova08}.  
They suggested the presence of three molecular gas components in the complex, which correspond to quiescent undisturbed primordial 
gas (i.e. $-$18 km s$^{-1}$ $< V_{lsr} < -$15 km s$^{-1}$ (red)), gas compressed by the shock from expanding S235 H\,{\sc ii} 
region (i.e. $-$21 km s$^{-1}$ $< V_{lsr} < -$18 km s$^{-1}$ (central)), and gas expulsion from the embedded young star clusters driven 
by the combined effect of the cluster stars (i.e. $-$25 km s$^{-1}$ $< V_{lsr} < -$21 km s$^{-1}$ (blue)). 
Furthermore, \citet{kirsanova14} utilized NH$_{3}$ line data
and found the densest molecular clumps associated with East~1, East~2, and Central subregions, mostly belong to the ``central" molecular component.

Figure~\ref{fig8} shows the $^{12}$CO and $^{13}$CO integrated gray scale maps, obtained by integrating over the velocity 
range between $-$23 and $-$18 km s$^{-1}$. 
Bolocam dust continuum emission contours are also overlaid on the $^{12}$CO integrated map to illustrate the 
physical association of dust emission with the complex. 
The integrated CO maps trace a region empty of molecular gas 
between two subregions, namely North and North-West. 
It is to be noted that the region empty of CO gas is exactly coincident with the broken part in the extinction map, 
where the radio continuum emission at 610 MHz is elongated (see Figures~\ref{fig1a}b and~\ref{fig2a}a). 
This region also appears to be associated with Br$\alpha$ emission 
which is surrounded by H$_{2}$ and/or PAH features (see Figure~\ref{fig4ua}). 
It suggests a close interaction between ionized and molecular gas in the complex. 
It seems that the ionizing photons have managed to escape in this direction. 
We propose that these features are best explained by the existence of a cavity 
between North and North-West subregions (also see Section~\ref{subsec:phyp}).

In Figure~\ref{fig9}, we show position-velocity (p-v) analysis of $^{12}$CO and $^{13}$CO gas. 
The p-v plots (right ascension-velocity and declination-velocity) of $^{12}$CO and $^{13}$CO gas have revealed 
noticeable velocity gradients and show 
essentially the same structure. However, $^{13}$CO data provide more information related to the velocity 
structure of the denser regions compared to $^{12}$CO emission (see declination-velocity plots in Figure~\ref{fig9}). 
The p-v plots of $^{12}$CO and $^{13}$CO gas reveal an almost semi-ring-like or C-like structure at larger scale (about 8 parsec extended). 
The detection of such morphology in the p-v plot is often interpreted as signature of an expanding shell \citep[e.g.][]{wilson05,arce11,dewangan15b}. 
We find an expansion velocity of the gas to be $\sim$3 km s$^{-1}$.  
Detailed discussion about this morphology can be found in Section~\ref{subsec:phyp}. 
Note that the S235 complex hosts the O9.5V type star, therefore the presence of the C-like structure in the 
complex can be interpreted as caused by the expanding H\,{\sc ii} region. 
Similar result was also observed in W42 star-forming region \citep[see Figure~8 in][]{dewangan15b}.
\subsubsection{Molecular outflows}
In addition to the C-like structure in the p-v plots, we have also found noticeable velocity gradients 
toward the East~1, East~2, Central~E, and North-West subregions, suggesting the presence of outflow activities in each of 
subregions. Following Paper~I, we know that the East~1, East~2, and Central~E subregions contain a cluster of YSOs. 
We have searched for outflows toward these subregions using the doppler shifted velocity components (red, green, and blue).
Due to the coarse beam of CO data (beam size $\sim$46$\arcsec$), we cannot pinpoint the exact exciting source of the outflows 
in each subregion. Considering this limitation, we have not shown the doppler shifted components (red, green, and blue) toward 
the East~2, Central~E, and North-West subregions. 
In Figure~\ref{fig10}, we present the noticeable receding, approaching, and rest gas components only in East~1 subregion. 
We have also marked the position of at least one probable driving source of the CO outflow (also see Section~\ref{subsec:h2out}). 
High resolution molecular line observations are necessary to examine better insight into the molecular outflows in the complex.
\subsection{NH$_{3}$ radial velocity and non-thermal velocity dispersion}
\label{subsec:amonia}
As mentioned in the introduction, in general, it is known that NH$_{3}$ data trace the densest regions of molecular cloud in a given star-forming region. 
The properties of densest gas traced by NH$_{3}$ line observations have not yet been studied 
with respect to the ionizing star in the S235 complex. 
The locations of NH$_{3}$ line observations are marked in Figure~\ref{fig2a}b, which were mainly observed in 
East~1, East~2, and Central~E subregions \citep[see][for more details]{kirsanova14}. 
In Figure~\ref{fig15}a, we show the variation of NH$_{3}$(1,1) radial velocity in each subregion as a function of distance from the location of the O9.5V star. 
The NH$_{3}$(1,1) radial velocities vary between $-$22 and $-$18.5 km~s$^{-1}$ in these three subregions. 
In general, there are noticeable velocity variations within East~1 ($-$21.5 to $-$19.0 km~s$^{-1}$), 
East~2 ($-$21.5 to $-$20.5 km~s$^{-1}$), and Central~E ($-$22 to $-$18.5 km~s$^{-1}$) subregions. 
These observational characteristics are also evident in CO data (see Figure~\ref{fig9}). 
\citet{balser11} reported the velocity of ionized gas to be about $-25.61(\pm0.12)$ km s$^{-1}$ in the S235 H\,{\sc ii} region 
using a hydrogen radio recombination line (H87-93$\alpha$).  
Therefore, the molecular gas is red-shifted with respect to the ionized gas in the S235 H\,{\sc ii} region. 
Interestingly,  we investigate that the radial velocity observed in Central~E subregion shows a linear 
trend as we move away from the location of a massive star, 
which is not seen in other two subregions. In each subregion, we have line of sight velocity components and 
it seems that the radial velocity of gas in Central~E subregion might be almost equal to the real velocity, 
however it is not the case for other two subregions. 
The observed trend could be interpreted as the direct influence of the expanding H\,{\sc ii} region excited by the O9.5V star.
Further discussion on this result is presented in Section~\ref{sec:disc}. 

We computed thermal sound speed ($a_{s}$), non-thermal velocity dispersion ($\sigma_{NT}$), and the ratio of thermal to 
non-thermal pressure ($P_{TNT} = {a_s^2}/{\sigma^2_{NT}}$). 
One can find more details about $P_{TNT}$ in the work of \citet{lada03}. 
The sound speed $a_{s}$ (= $(k T_{kin}/\mu m_{H})^{1/2}$) can be estimated with the knowledge of 
gas kinetic temperature (T$_{kin}$) and $\mu$=2.37 (approximately 70\% H and 28\% He by mass). 
The non-thermal velocity dispersion is given by:
\begin{equation}
\sigma_{\rm NT} = \sqrt{\frac{\Delta V^2}{8\ln 2}-\frac{k T_{kin}}{17 m_H}} = \sqrt{\frac{\Delta V^2}{8\ln 2}-\sigma_{\rm T}^{2}} ,
\label{sigmanonthermal}
\end{equation}
where $\Delta V$ is the measured FWHM linewidth of the observed NH$_{3}$ spectra, 
$\sigma_{\rm T}$ (= $(k T_{kin}/17 m_H)^{1/2}$) is the thermal broadening for NH$_{3}$ at T$_{kin}$ \citep[e.g.][]{dunham11}, 
and $m_H$ is the mass of hydrogen atom. The computed values of $a_{s}$, $\sigma_{NT}$, and $P_{TNT}$ are given in Table~\ref{tab3}. 
The value of $P_{TNT}$ is computed for each of the observed positions.  
In each subregion, the variation of $P_{TNT}$ as a function of distance from the location of a massive star is shown in Figure~\ref{fig15}b. 
We find an average $P_{TNT}$ value of 0.13, 0.21, and 0.22 for East~1, East~2, and Central~E subregions, respectively. 
It suggests that non-thermal pressure is dominant in these densest subregions. 
The lower values of $P_{TNT}$ might be due to contributions from the bipolar outflows, as investigated in previous section. 
The average values of Mach numbers ($\sigma_{\rm NT}$/$a_{s}$) estimated for East~1, East~2, and Central~E 
subregions are 2.9, 2.3, and 2.9, respectively which indicate that these subregions are supersonic. 
\subsection{Feedback of a massive star}
\label{subsec:feed}
In this section, we derive the various pressure components (pressure of an H\,{\sc ii} region $(P_{HII})$, radiation pressure (P$_{rad}$), 
and stellar wind ram pressure (P$_{wind}$)) driven by a massive star to study its feedback in the vicinity.

It is found that the massive star is approximately located at the center of the sphere-like shell. 
The radio continuum images presented in this work can be used to estimate the number of Lyman continuum photons (N$_{uv}$).
The expression of N$_{uv}$ is given by \citep{matsakis76}:
\begin{equation}
N_{uv} (s^{-1}) = 7.5\, \times\, 10^{46}\, \left(\frac{S_{\nu}}{Jy}\right)\left(\frac{D}{kpc}\right)^{2} 
\left(\frac{T_{e}}{10^{4}K}\right)^{-0.45} \\ \times\,\left(\frac{\nu}{GHz}\right)^{0.1}
\end{equation}
\noindent where S$_{\nu}$ is the measured total flux density in Jy, D is the distance in kpc, 
T$_{e}$ is the electron temperature, and $\nu$ is the frequency in GHz. 
The calculation of N$_{uv}$ is performed for both the radio frequencies (0.61 GHz and 1.4 GHz) separately. 
Substituting D = 1.8 kpc, T$_{e}$ = 10000 K, and S$_{1.4}$ = 1.92 Jy, we compute N$_{uv}$ = 4.9 $\times$ 10$^{47}$ s$^{-1}$.
Similarly, we find N$_{uv}$ = 8.2 $\times$ 10$^{47}$ s$^{-1}$ for D, T$_{e}$, and S$_{0.61}$ = 3.49 Jy. 
The estimates of ionizing photon flux values at different frequencies correspond to a single ionizing star of O9.5V spectral type \citep{martins05}, 
which is also in agreement with the previously reported spectral type of the ionizing source of the complex \citep[e.g.][]{georgelin73}.  

The different pressure components ($P_{HII}$, P$_{rad}$, and P$_{wind}$; as mentioned above) are defined as below \citep[e.g.][]{bressert12}:
\begin{equation}
P_{HII} = \mu m_{H} c_{s}^2\, \left(\sqrt{3N_{uv}\over 4\pi\,\alpha_{B}\, D_{s}^3}\right);\\ 
\end{equation}
\begin{equation}
P_{rad} = L_{bol}/ 4\pi c D_{s}^2; \\ 
\end{equation}
\begin{equation}
P_{wind} = \dot{M}_{w} V_{w} / 4 \pi D_{s}^2; \\
\end{equation}
In the equations above, N$_{uv}$, $\mu$, and m$_{H}$ are defined earlier, the radiative recombination 
coefficient is ``$\alpha_{B}$'' \citep[=  2.6 $\times$ 10$^{-13}$ $\times$ (10$^{4}$ K/T$_{e}$)$^{0.7}$ cm$^{3}$ s$^{-1}$; see][]{kwan97}, 
c$_{s}$ is the sound speed in the photo-ionized region (= 10 km s$^{-1}$), $\dot{M}_{w}$ is the mass-loss rate, 
V$_{w}$ is the wind velocity of the ionizing source,  L$_{bol}$ is the bolometric luminosity of the complex, and 
D$_{s}$ is the projected distance from the location of the O9.5V type star to the subregions 
where the pressure components are estimated. 
One can infer from equations 5, 6, and 7 that the pressures, P$_{rad}$ and P$_{wind}$, 
scale as $D_{s}^{-2}$ while P$_{HII}$ scales as $D_{s}^{-3/2}$.\\ 

\begin{figure*}
\epsscale{1.0}
\plotone{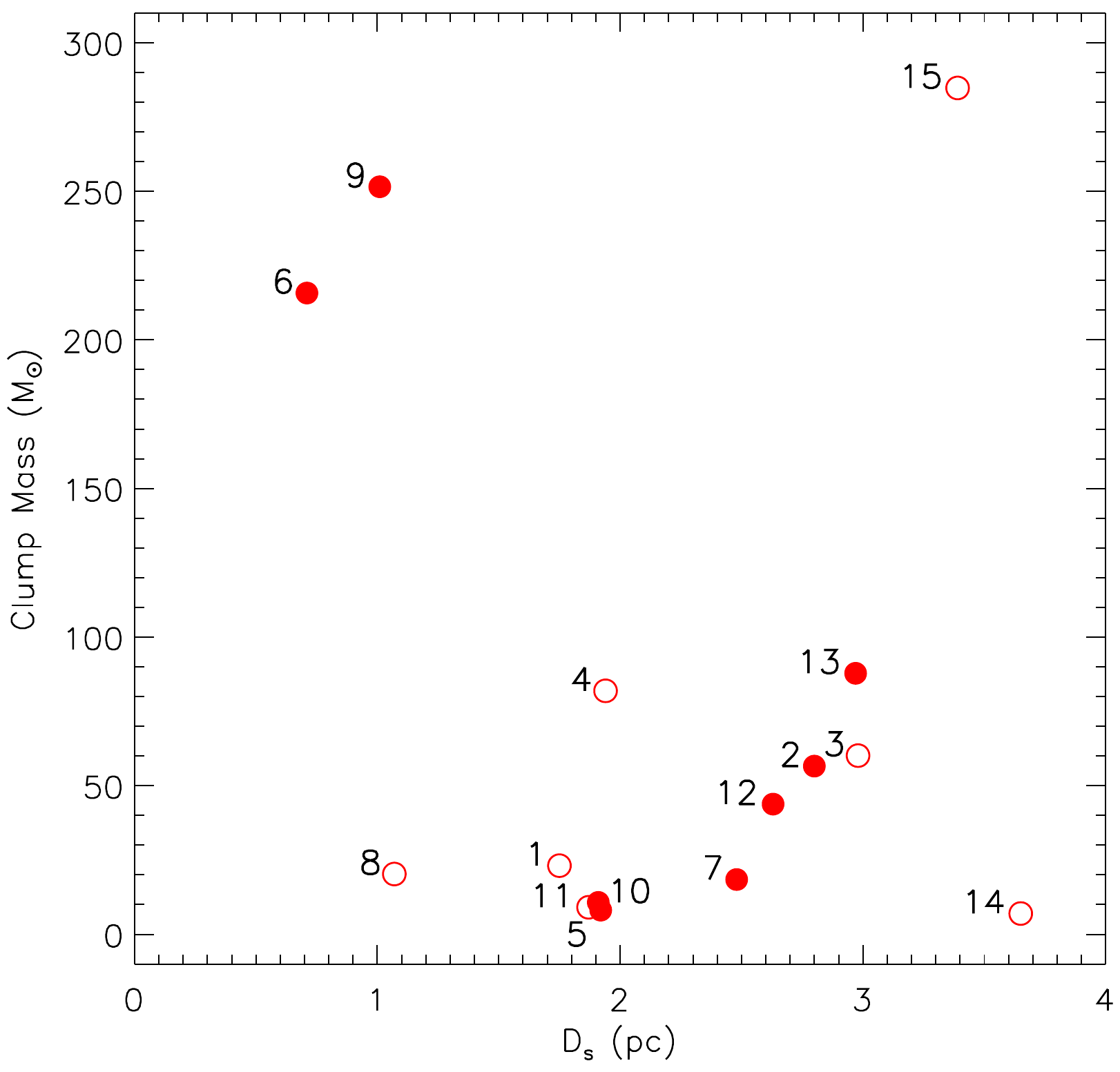}
\caption{\scriptsize The distribution of clump masses (estimated using Bolocam 1.1 mm data) with respect to the location of an O9.5V star (D$_{s}$). 
Filled circles show the clumps distributed near the edges of the sphere-like shell 
(i.e. Central~E, East~2, North, and North-West subregions; see Figure~\ref{fig2a} and Table~\ref{tab1}).}
\label{fig3a}
\end{figure*}

\begin{figure*}
\epsscale{1.0}
\plotone{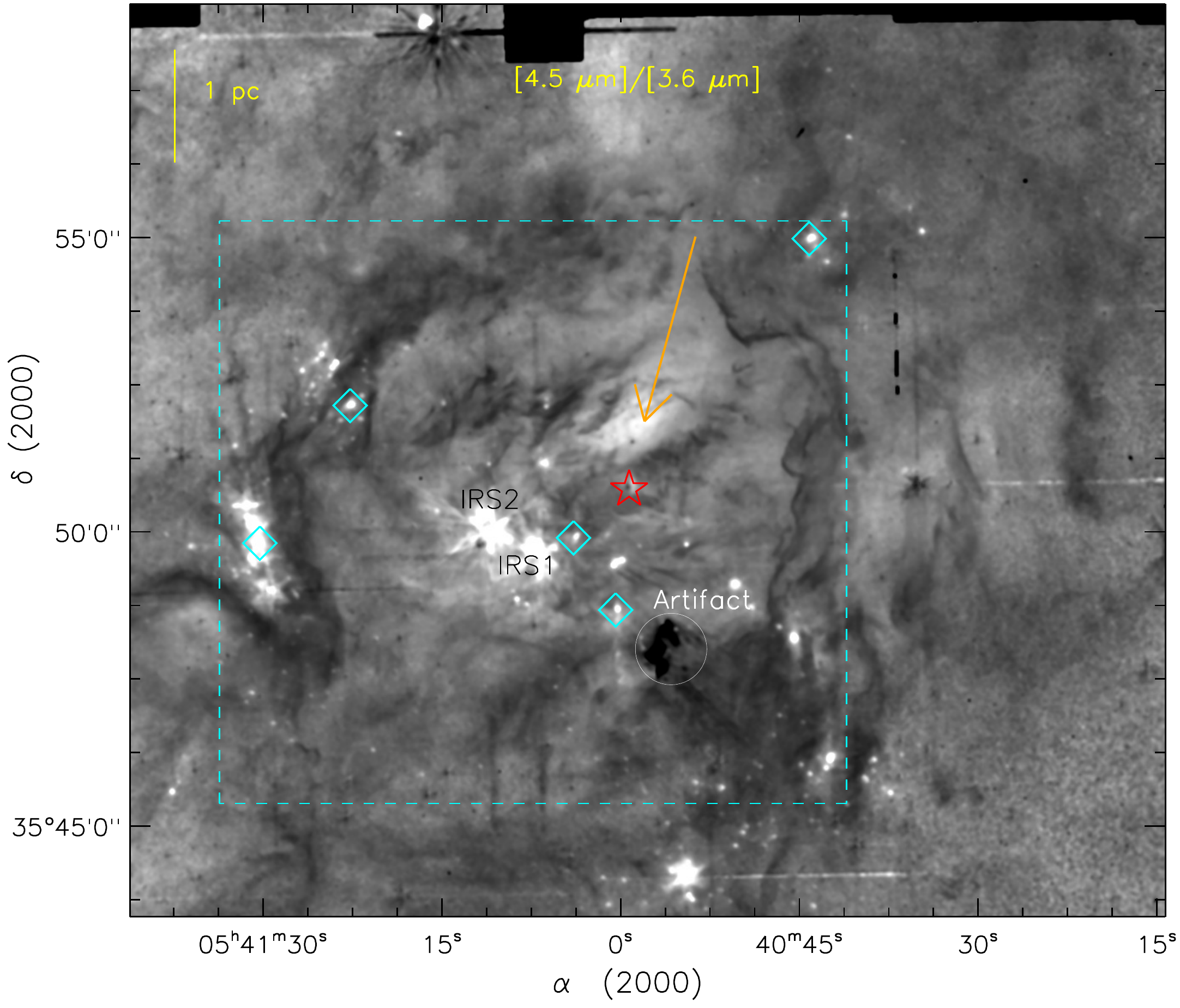}
\caption{\scriptsize {\it Spitzer}-IRAC ratio map of 4.5 $\mu$m/3.6 $\mu$m emission (similar area as shown in Figure~\ref{fig1a}). 
The area enclosed in the white circle shows artifact present in the map. The location of an O9.5V star (BD+35$^{\degr}$1201) is 
marked with a star. The diamond symbols indicate the possible outflow activities (see text for more details). 
An arrow indicates the bright emission in the map, which probably traces the Br$\alpha$ emission. 
Note that this bright emission is surrounded by black regions.
The dashed box is shown as a zoomed-in view in Figure~\ref{fig5}. 
Two bright sources (i.e. IRS~1 and IRS~2) have sidelobes therefore the features 
around these sources in the ratio map can not be considered for scientific interpretation. 
The ratio map is subjected to median filtering with a width of 5 pixels and smoothened by 
4 pixel $\times$ 4 pixel using a boxcar algorithm.}
\label{fig4ua}
\end{figure*}

\begin{figure*}
\epsscale{1.0}
\plotone{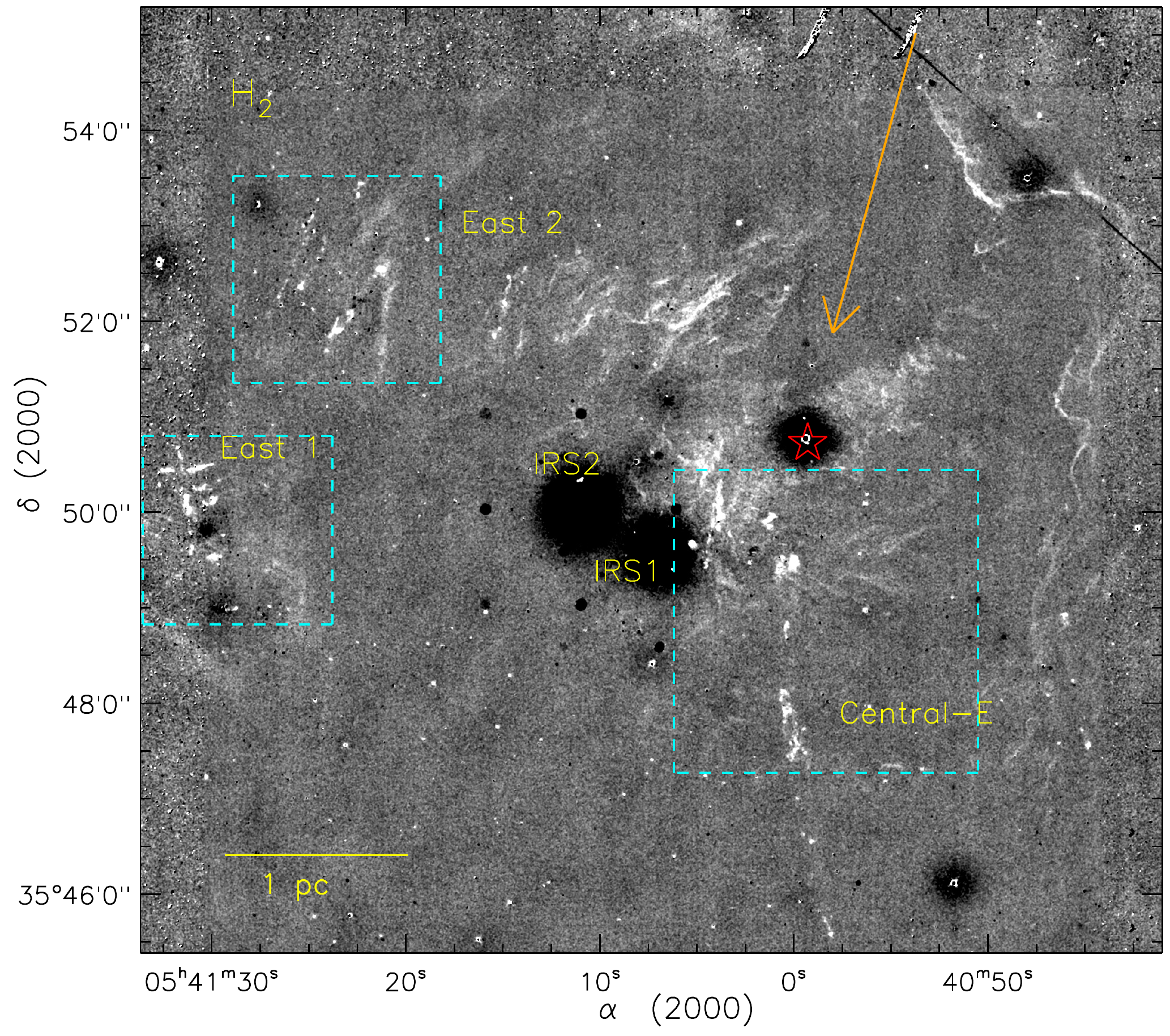}
\caption{\scriptsize CFHT continuum-subtracted H$_{2}$ map (gray-scale) at 2.12 $\mu$m of the 
S235 complex (size of the region $\sim$10$\farcm$7 $\times$ 9$\farcm$9 or $\sim$5.6 pc $\times$ 5.2 pc), 
as shown by a dashed box in Figure~\ref{fig4ua}. 
The location of an O9.5V star (BD+35$^{\degr}$1201) is marked with a star symbol. 
An arrow in orange color is similar to the one shown in Figure~\ref{fig4ua}. 
The H$_{2}$ map is subjected to median filtering with a 
width of 3 pixels and smoothened by 3 pixel $\times$ 3 pixel using a boxcar algorithm to enhance the faint features. 
The dashed boxes are shown as a zoomed-in view in Figure~\ref{fig6}. }
\label{fig5}
\end{figure*}

\begin{figure*}
\epsscale{1.0}
\plotone{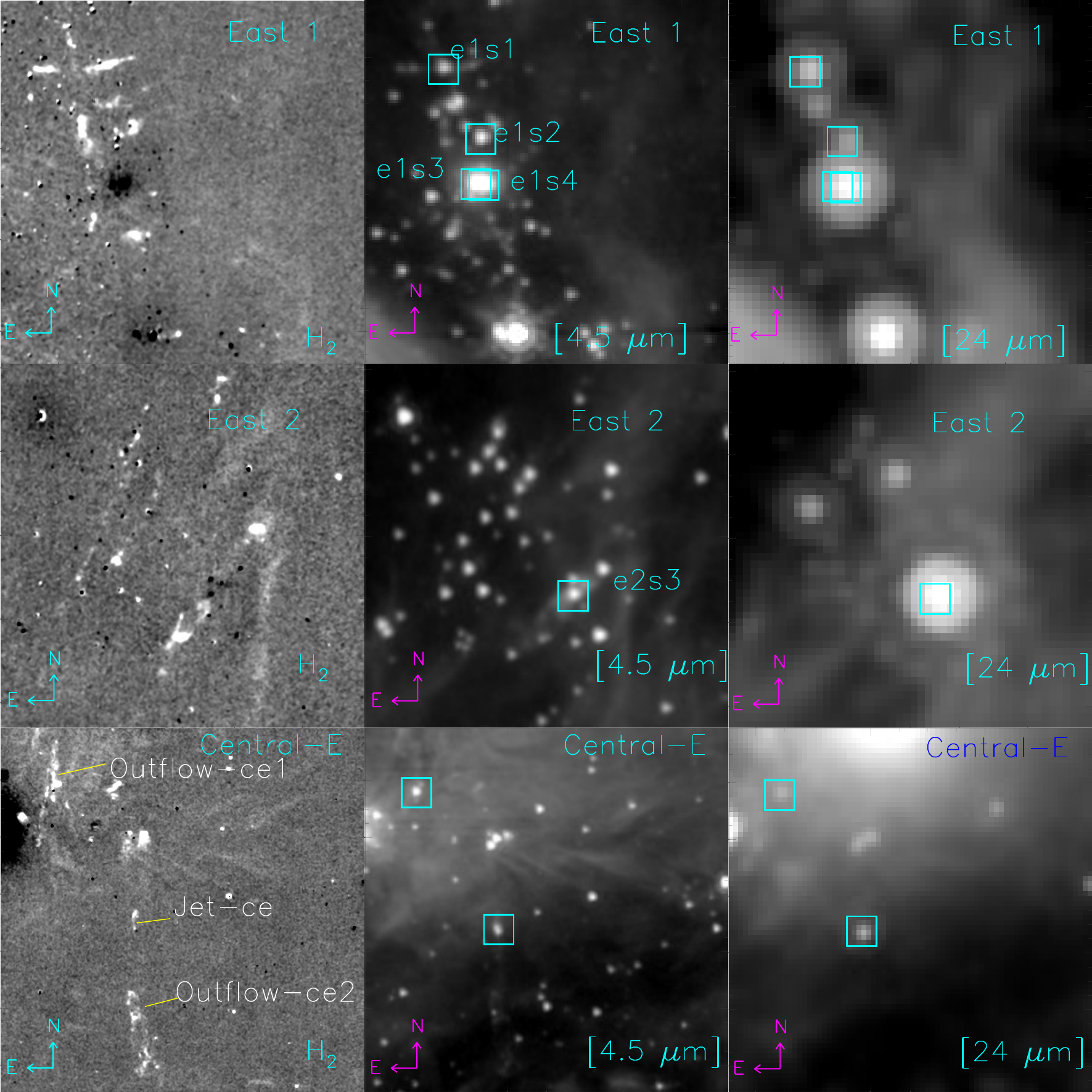}
\caption{\scriptsize A comparison of H$_{2}$, 4.5 $\mu$m, and 24 $\mu$m images toward three 
subregions (East~1, East~2, and Central-E) in the S235 complex 
(see dashed boxes in Figure~\ref{fig5}). 
Left panel: CFHT continuum-subtracted H$_{2}$ maps (log-scale) at 2.12 $\mu$m. 
Jet-like features and outflows are marked in the figures. 
The H$_{2}$ maps are subjected to median filtering with a 
width of 3 pixels and smoothened by 3 pixel $\times$ 3 pixel using a boxcar algorithm to enhance the faint features. 
Middle panel: IRAC 4.5 $\mu$m images (log-scale). 
Right panel: MIPS 24  $\mu$m images (log-scale). The positions of the probable driving sources of outflows are shown by open 
squares in both the middle and right panels. 
Sources are also labeled in East~1 and East~2 subregions in the middle panel.}
\label{fig6}
\end{figure*}

\begin{figure*}
\epsscale{0.70}
\plotone{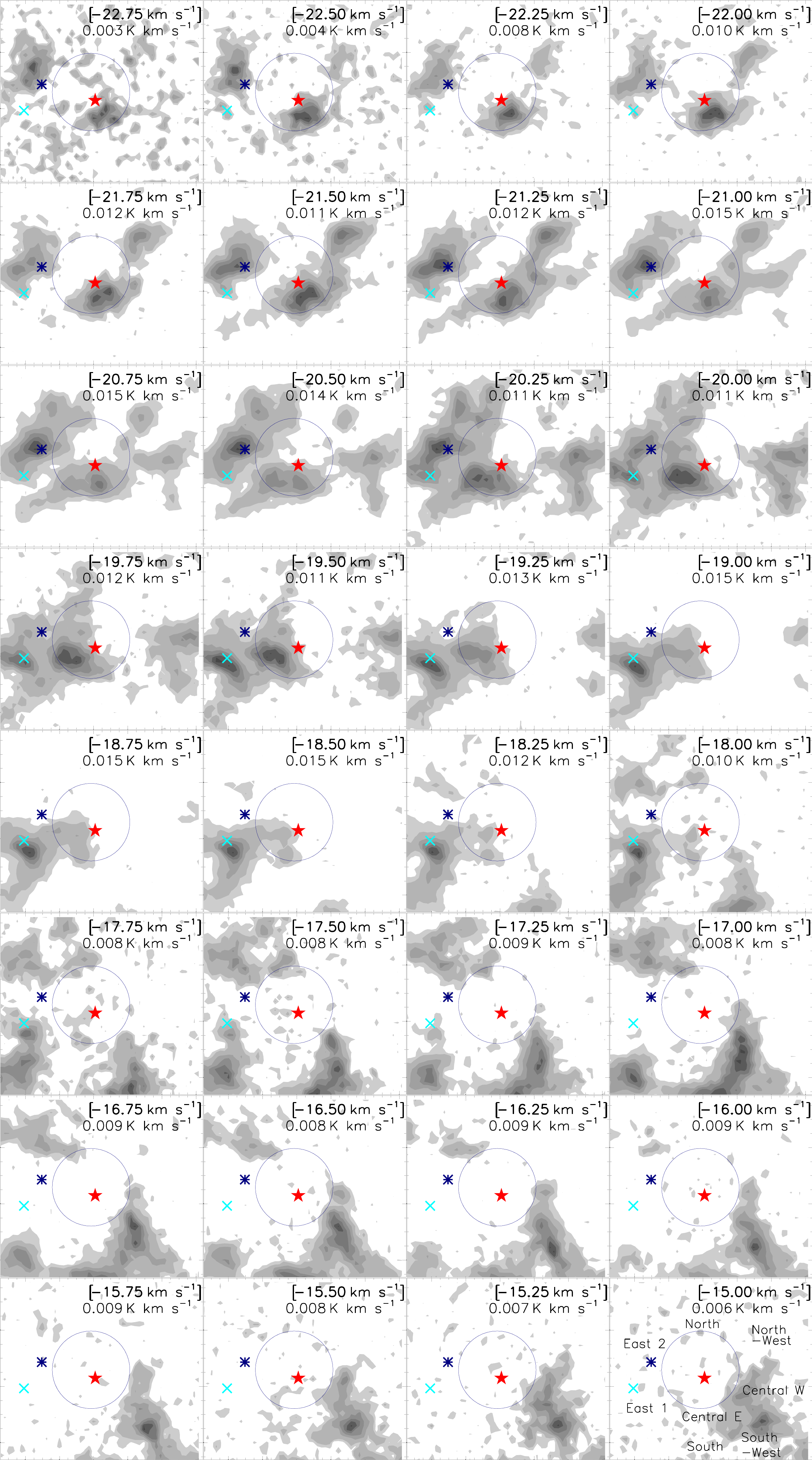}
\caption{\scriptsize The $^{13}$CO(J =1$-$0) velocity channel contour maps of the S235 complex. 
The velocity value (in km s$^{-1}$) is indicated in each panel. 
The contour levels are 10, 20, 40, 55, 70, 85, and 98\% of the peak value (in K km s$^{-1}$), which is also given in each panel.  
The location of an O9.5V star is marked by a filled star. 
Other marked symbols and labels are similar to those shown in Figure~\ref{fig2a}b.}
\label{fig7}
\end{figure*}

Note that the highlighted subregions are not located at the same projected distance from 
the location of a massive star (see Figures~\ref{fig2a} and~\ref{fig3a}). 
Therefore, we compute the pressure components driven by a massive star (i.e., $P_{HII}$,  P$_{rad}$, and P$_{wind}$) 
at D$_{s}$ = 3.4 pc (East~1), 3.0 pc (East~2), 1.5 pc (Central~E), 2.6 pc (North), 2.7 pc (North-West), 1.9 pc (Central~W), and 3.0 pc (South-West) (see Figures~\ref{fig2a} and~\ref{fig3a}). 

\citet{evans81a} estimated a total gas mass of the S235 cloud to be about 3000 $M_\odot$ in a radius of 3 pc. 
\citet{nordh84} computed a total luminosity of S235 of $\sim$8 $\times$ 10$^{4}$ L$_{\odot}$.
We use $M_{cloud}$ $\approx$ 3000 $M_\odot$, R$_{c}$ $\approx$ 3.0 pc, $L_{bol}$ $\approx$ 8 $\times$ 10$^{4}$ L$_{\odot}$, 
$\dot{M}_{w}$ $\approx$ 1.58 $\times$ 10$^{-9}$ M$_{\odot}$ yr$^{-1}$ \citep[for O9.5V star;][]{marcolino09}, 
V$_{w}$ $\approx$ 1500 km s$^{-1}$ \citep[for O9.5V star;][]{marcolino09} in the above equations to compute the pressure contributions 
driven by a massive star on different subregions (see the list in Table~\ref{tab4}). 
The error associated with each pressure component related to each subregion is also given in Table~\ref{tab4}.  
In the S235 complex, we find that the pressure of the H\,{\sc ii} region exceeds the radiation pressure and the 
stellar wind pressure (see Table~\ref{tab4}).  
The total pressure (P$_{total}$ = P$_{HII}$ + $P_{rad}$ + P$_{wind}$) driven by a massive star on each subregion is also tabulated in 
Table~\ref{tab4}, which is found to be $\sim$10$^{-10}$ dynes cm$^{-2}$ for each subregion.

Additionally, pressure exerted by the self-gravity of the surrounding molecular gas is also estimated using the relation \citep[e.g.][]{harper09}:
\begin{equation}
P_{cloud} \approx\pi G\Sigma^2\\
\end{equation}
\noindent where $\Sigma$ $(= M_{cloud}/\pi R_{c}^2)$ is the mean mass surface density of the cloud, 
M$_{cloud}$ is the mass of the molecular gas, and R$_{c}$ is the radius of the molecular region. 
The value of $P_{cloud}$ for the entire cloud is estimated to be $\sim$(1.0$\pm$0.5) $\times$ 10$^{-10}$ dynes cm$^{-2}$. 
The values of $P_{cloud}$ associated with different subregions are also found to be $\sim$0.1--3.0 $\times$10$^{-10}$ dynes 
cm$^{-2}$ (see Table~\ref{tab1} and Figure~\ref{fig2a}a). 
In subregions, we find P$_{total}$ $\gtrsim$ $P_{cloud}$ (see Table~\ref{tab4}). 
This argument is still valid if we put all the subregions at the same projected distance 
(i.e. D$_{s}$ = 3.0 pc; see pressure calculations related to East~2 in Table~\ref{tab4}). 
Additionally, the $P_{cloud}$ is relatively higher than the pressure associated with a typical cool molecular cloud 
($\sim$10$^{-11}$ -- 10$^{-12}$ dynes cm$^{-2}$ for a temperature $\sim$20 K and the particle density $\sim$10$^{3}$--10$^{4}$ cm$^{-3}$) 
\citep[see Table 7.3 of][]{dyson80}. 
Considering the pressure value associated with a typical cool molecular cloud, 
we infer that an additional physical process has been acted to compress the surrounding molecular gas 
to enhance the pressure in the complex. 
Consequently, this process may stimulate the initial collapse and fragmentation of the extended molecular cloud.  
The pressure calculations indicate that the photoionized gas can be considered as the 
important contributor for the feedback mechanism in the S235 complex.
\subsection{Young stellar objects in S235}
\subsubsection{Identification of young stellar objects}
\label{subsec:phot1}
A population of YSOs can be identified using NIR and MIR color schemes as utilized by many authors 
\citep[e.g.][]{allen04,lada06,gutermuth09}. In Paper~I, we identified YSOs using only IRAC data. 
In the present work, we also employed the UKIDSS-GPS NIR data, in combination with IRAC data for finding more deeply embedded 
and faint sources. Note that the UKIDSS-GPS NIR data are three magnitudes deeper than 2MASS. 
Here, we describe the procedure of YSOs identification and classification using photometric IRAC and GPS data.\\

1. We selected sources having detections in all four {\it Spitzer}-IRAC 
bands, which were taken from Paper~I. 
We found 131 YSOs (52 Class~I and 79 Class~II), 4 Class~III, 133 photospheres, 
and 38 contaminants in our selected region shown in Figure~\ref{fig1a}a. 
The details of YSOs classification can be found in Paper~I. Figure~\ref{fig16}a displays the different selected sources. 
In Figure~\ref{fig16}a, we show Class~I YSOs (red circles), Class~II YSOs (open blue triangles), 
Class~III YSOs (black squares), photospheric emissions (gray dots), and 
PAH-emission-contaminated apertures (magenta multiplication symbols). \\ 

2. Then, we considered sources lacking detections in two longer 
wavelengths of IRAC bands (5.8 and 8.0 $\mu$m). 
For such sources, GPS-IRAC (H, K, 3.6, and 4.5 $\mu$m) classification scheme was utilized, as described in details by 
\citet{gutermuth09}. In this method, the dereddened color-color space ([K$-$[3.6]]$_{0}$ and [[3.6]$-$[4.5]]$_{0}$) was used to identify infrared-excess sources.
These dereddened colors were computed using the color excess ratios given in \citet{flaherty07}. 
This scheme also offered to identify possible dim extragalactic contaminants from YSOs with additional conditions (i.e., [3.6]$_{0}$ $<$ 15 mag for Class~I and [3.6]$_{0}$ $<$ 14.5 mag for Class~II). 
We used the observed color and the reddening law \citep[from][]{flaherty07} to compute the dereddened 3.6 $\mu$m magnitudes. 
In the end, 16 Class~I and 133 Class~II YSOs are obtained using GPS-IRAC data (see Figure~\ref{fig16}b). \\

3. NIR color-color diagram (H$-$K vs J$-$H) is a very useful tool to select infrared-excess sources. 
We applied this scheme for sources, having detections in all the three JHK bands. 
Figure~\ref{fig16}c shows a NIR color-color diagram of such sources. 
The reddening lines are drawn using the extinction law of \citet{indebetouw05}. 
The color-color diagram is divided into three different regions, namely ``I'', ``II'', and ``III''.  One can find more details about 
NIR YSOs classification in \citet{sugitani02} and \citet{dewangan12}.  
The sources distributed in the ``I'' region represent the likely Class~I objects (protostellar objects). 
T Tauri-like sources (Class~II objects) are identified within the ``II'' region along the T Tauri locus with large NIR excess. 
The sources that fall between the reddening bands of the main-sequence and giant stars are located in the ``III'' region. 
With this method, we obtain 5 Class~I and 96 Class~II sources (see Figure~\ref{fig16}c). \\ 

4. Finally, we considered sources (significant in number) that have detections only in the  H and K bands. 
In order to further identify infrared excess sources from such sample, we utilized a color-magnitude (H$-$K/K) diagram (see Figure~\ref{fig16}d). 
The diagram allows to select red sources having H$-$K $>$ 1.04. 
The color-magnitude analysis of the nearby control field allowed us to infer this color criterion. 
This condition provides us 54 additional deeply embedded infrared excess sources.\\

Recently, \citet{chavarria14} also presented photometry of the entire extended star-forming S235 region (including S235AB region) using NIR and {\it Spitzer} data. 
These authors obtained K-band observations using the 2.1-m telescope located at Kitt Peak National Observatory. 
Additionally, they observed J and H bands using the 6.5-m Multiple Mirror Telescope (MMT) telescope located at Fred Lawrence Whipple Observatory. 
They identified 690 YSOs in the entire S235 complex. However, they did not separate the contaminations in their analysis. 
Additionally, they did not provide the positions of their identified YSOs. Therefore, a comparison of our selected YSOs to the sources identified by 
\citet{chavarria14} cannot be  thoroughly performed in this work (see Figure~\ref{fig17} in this work and Figure B2 in \citet{chavarria14}). 
In our YSOs analysis, we have carefully separated the contaminations from YSO populations. 
As a result of the four schemes explained above, a total of 435 YSOs are yielded in the complex (S235AB region is not included). 
The positions of all YSOs are shown in Figure~\ref{fig17}. 

\subsubsection{Spatial distribution of YSOs}
\label{subsec:surfden}
In recent years, the surface density of young stellar populations in star-forming regions has been adopted to identify 
and to study the young stellar clusters \citep[e.g.][]{gutermuth09,bressert10}. In Paper~I, the surface density map of YSOs was 
computed using the nearest-neighbour (NN) technique \citep[also see][for more details]{gutermuth09,bressert10}. 
Here, we also followed the same procedure and generated the surface density map of all the selected 435 YSOs.
The map was created using a 5$\arcsec$ grid and 6 NN at a distance of 1.8 kpc. 
Figure~\ref{fig17} shows the resultant surface density contours of YSOs overlaid on the extinction map.
The contour levels are shown at 10, 20, 40, and 70 YSOs/pc$^{2}$, increasing from the outer to the inner regions. 
The figure clearly illustrates the spatial correlation between YSOs surface density and extinction. 

The surface density analysis allows to trace the individual groups or clusters of YSOs. 
Now, we want to identify the clustered populations from distributed sources. 
However, there is no unique method to find a cutoff distance for tracing these two populations.  
Statistical analysis, such as empirical cumulative distribution (ECD) of YSOs as a 
function of NN distance, is often studied to infer the clustered YSO populations in 
star-forming regions \citep[see][for more details]{chavarria08,gutermuth09,dewangan11}. 
In the ECD analysis, a cutoff length (also referred as the distance of inflection d$_{c}$) is chosen, to delineate the low-density/distributed populations. 
For the present case, we selected a cutoff distance of d$_{c}$ $\sim$52$\arcsec$ (0.45 pc at a distance of 1.8 kpc), 
which separates the cluster members within the contour level of 10 YSOs/pc$^{2}$. 
Such cutoff distance has been determined for different star-forming regions, 
such as the Diamond Ring region in the Cygnus-X star-forming complex \citep[d$_{c}$ $\sim$61$\arcsec$ or 0.43 pc at 1450 pc;][]{beerer10}, 
the Cygnus-OB2 region in the Cygnus-X complex \citep[d$_{c}$ $\sim$72$\arcsec$ or 0.51 pc;][]{guarcello13}, and the W5 star-forming region 
\citep[d$_{c}$ $\sim$33$\arcsec$ or 0.32 pc at 2 kpc;][]{chavarria14}. 
The comparison of d$_{c}$ values suggests that W5 is more densely populated with YSOs compared to Cygnus-X. 
In general, the value of d$_{c}$ offers to infer the most densely YSOs populated star-forming regions. 
The ECD analysis yielded a clustered fraction of about 59\% YSOs (i.e. 258 from a total of 435 YSOs). 
The YSO clusters are mainly distributed in East~1, East~2, Central~E, Central~W, North-West, South, and South-West subregions. 
In Central~E and East~1 subregions, we found the highest level of surface density contours. 
North subregion is not associated with any YSO cluster; however, it contains a few Class~II YSOs. 
The clustered and scattered YSO locations are identified in Figure~\ref{fig17} by filled and open symbols, respectively.

We also constructed separately the surface density contour maps of Class~I 
and Class~II YSOs, using the same algorithm as explained above.  
The surface density contour maps of Class~I and Class~II YSOs are shown in Figure~\ref{fig18}a. 
The surface density of Class~I YSOs lies between 0.04 and 42.75 YSOs/pc$^{2}$ with a dispersion ($\sigma$) of 2.64 
and is shown by contour levels of 5, 8, 10, and 20 YSOs/pc$^{2}$. 
Similarly, the surface density of Class~II YSOs is found to lie between 0.17 and 174 YSOs/pc$^{2}$, with a dispersion of 6.62 
and is drawn at contour levels of 10, 20, and 40 YSOs/pc$^{2}$. 
All the Class~I and Class~II YSO clusters are exclusively found in the regions with A$_{V} > 8$ mag.
The Class~I YSO clusters are traced in East~1, East~2, Central~E, South, and South-West subregions (see Figure~\ref{fig18}a). 
\section{Discussion}
\label{sec:disc}
\subsection{Expanding H\,{\sc ii} region}
\label{subsec:phyp}
One of the important results of the present work is the identification of 
an almost semi-ring-like or C-like structure in the p-v plots, 
which is similar to the one observed in Orion nebula \citep[see Figure~10b in][]{wilson05} and to that in W42 \citep[see Figure~8 in][]{dewangan15b}. 
\citet{arce11} also studied such observed structures in Perseus molecular 
cloud along with the modeling of expanding bubbles in a turbulent medium. 
These authors suggested that the semi-ring-like or C-like structure in the p-v plots is the signature of an expanding shell.
They also mentioned that the ring-like morphology can be seen in the p-v plot when the powerful ionizing source is 
situated at the center of the region.
Additionally, the NH$_{3}$ radial velocity observed in Central~E subregion exhibits a linear trend as one moves away from the location of a massive star.
The analysis of pressure contributions driven by a massive star (i.e., P$_{HII}$, P$_{rad}$, and P$_{wind}$) 
indicates that the P$_{HII}$ value is higher than the other pressure components. 
Therefore, it seems that the C-like structure in the p-v plots corresponds to the expanding H\,{\sc ii} region excited by the O9.5V star at 
the center of the sphere-like shell. Negative feedback of the O9.5V star is also evident by the presence of a 
cavity (i.e. empty of molecular CO gas region) between North and North-West subregions. 
Previously, \citet{evans81a} also suggested that the S235 molecular cloud is heated 
by the exciting star. All these observed features suggest the impact of ionizing photons on 
the low density regions in the complex, through which the ionized gas might have escaped. 
\subsection{Morphology and signposts of star formation}
\label{subsec:signp}
The NIR extinction map of the S235 complex provides a more complete picture of the spatial distribution of embedded subregions.
There are eight subregions which are well correlated with the locations of the dust clumps, CO gas, and YSOs.
Five of these subregions (i.e. Central~E, East~2, North, North-West, and Central~W) appear to be nearly 
regularly spaced along the sphere-like shell surrounding the ionized emission. \citet{deharveng03} also reported similar configuration in Sh~104 
region that was classified as a site of triggered star formation.
Furthermore, three subregions (i.e. East~1, South, and South-West) in the S235 complex are spatially away from the sphere-like shell. 
The presence of H$_{2}$ and PAH emissions at the boundaries of the sphere-like shell indicates the presence 
of PDR surrounding the ionized emission (see Figure~\ref{fig4ua}). 
In the present work, we investigated more number of YSOs using GPS, in addition to IRAC data, compared to Paper~I.  
The YSO clusters are found in all the subregions except North, where only Class~II YSOs are seen without any clustering.
Additionally, the clusters of Class~I YSOs are associated with Central~E, East~1, East~2, South, 
and South-West subregions. 
The molecular data also revealed outflow signatures in  the Central~E, East~1, East~2, and North-West subregions. 
We find the probable outflow driving candidate (i.e. Class~I YSO) in each subregion.  
All together, the early stages of star formation activity are evident in all the eight subregions.

\begin{figure*}
\epsscale{0.73}
\plotone{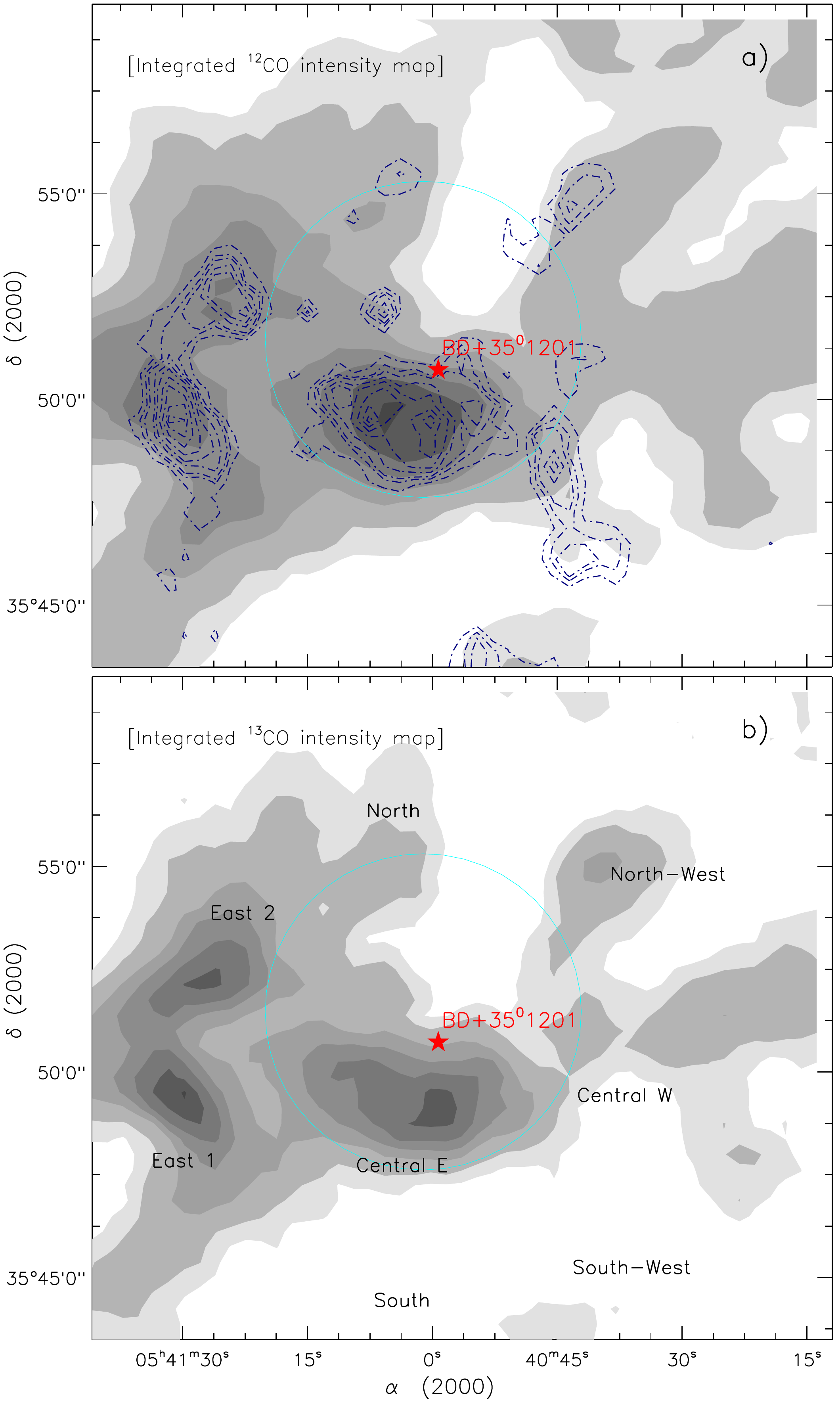}
\caption{\scriptsize The contour maps of integrated $^{12}$CO and $^{13}$CO emissions in the velocity range of $-$23 to $-$18 km s$^{-1}$. 
The contour levels are 10, 20, 40, 55, 70, 85, and 98\% of the peak values 
(i.e. 11.566 K km s$^{-1}$ ($^{12}$CO) and 31.719 K km s$^{-1}$ ($^{13}$CO)).  
In the top panel, Bolocam 1.1 mm dust continuum 
emission contours are shown by dashed navy contours with similar levels to those as shown in Figure~\ref{fig2a}b. 
In both the panels, other marked symbols and labels are similar to those shown in Figure~\ref{fig1a}.}
\label{fig8}
\end{figure*}

\begin{figure*}
\epsscale{1.0}
\plotone{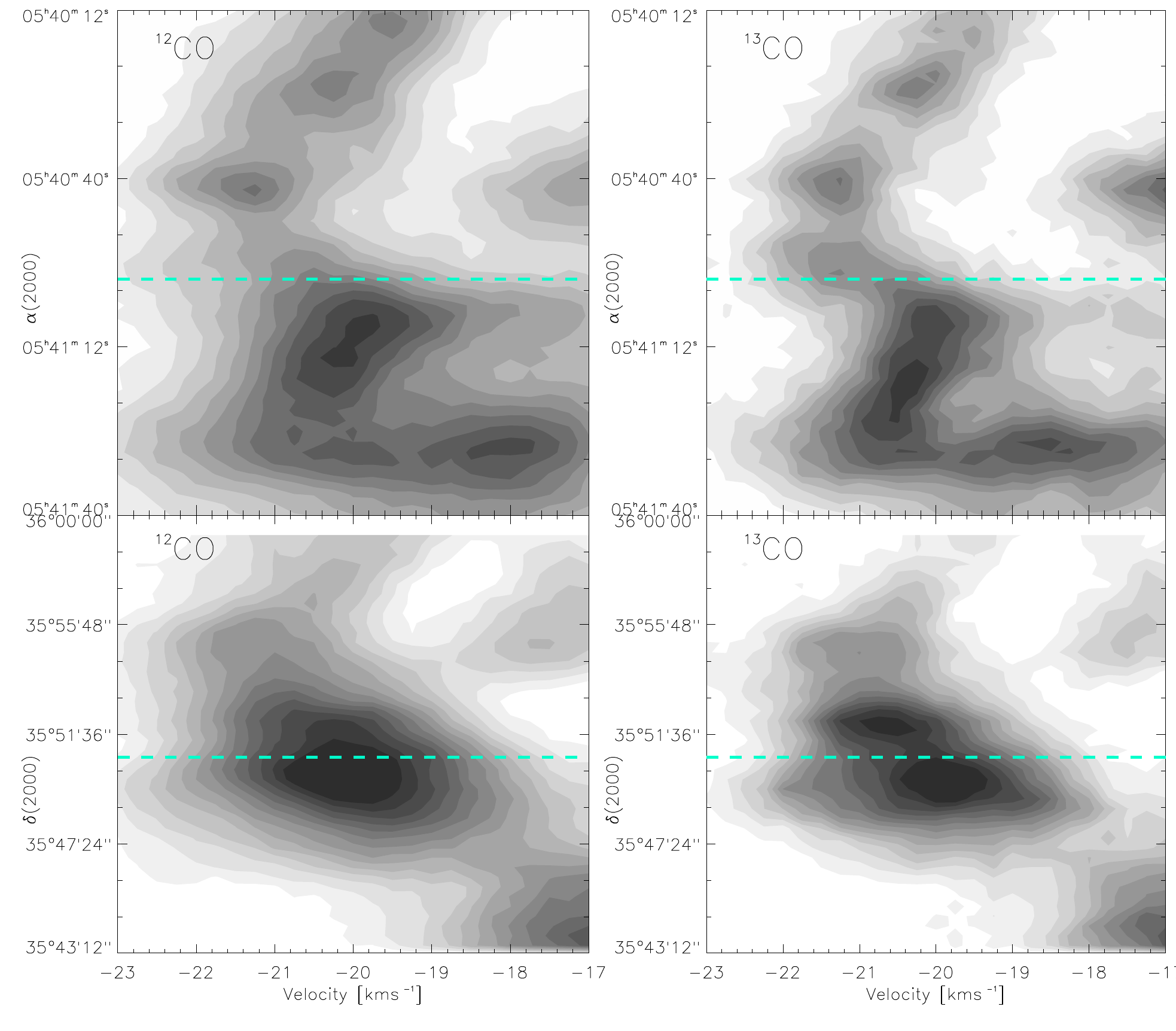}
\caption{\scriptsize The position-velocity diagrams of $^{12}$CO (left) and $^{13}$CO (right).
{\bf Top Left:} Right Ascension-velocity map of $^{12}$CO. 
The $^{12}$CO emission is integrated over the declination range from +35$\degr$43$\arcmin$30$\arcsec$ (35$\degr$.725) to 
+35$\degr$59$\arcmin$13$\farcs$2 (35$\degr$.987). 
The lowest gray-scale level corresponds to $30\sigma$ 
 (where $1\sigma$ = 0.01053 K degree), with successive levels increasing in steps of $30\sigma$. 
 {\bf Bottom Left:} Declination-velocity map of $^{12}$CO. The $^{12}$CO emission is integrated over the 
 right ascension range from 05$^{h}$41$^{m}$40$^{s}$.8 (85$\degr$.42) to 05$^{h}$40$^{m}$12$^{s}$ (85$\degr$.05). 
 The lowest gray-scale level corresponds to $30\sigma$ 
 (where $1\sigma$ = 0.011258 K degree), with successive levels increasing in steps of $30\sigma$. 
{\bf Top Right:} Right Ascension-velocity map of $^{13}$CO. 
The $^{13}$CO emission is integrated over the declination range from +35$\degr$43$\arcmin$30$\arcsec$ (35$\degr$.725) to 
+35$\degr$59$\arcmin$13$\farcs$2 (35$\degr$.987). The lowest gray-scale level corresponds to $15\sigma$ 
 (where $1\sigma$ = 0.005265 K degree), with successive levels increasing in steps of $15\sigma$. 
{\bf Bottom Right:} Declination-velocity map of $^{13}$CO. 
The $^{13}$CO emission is integrated over the right ascension range from 05$^{h}$41$^{m}$40$^{s}$.8 (85$\degr$.42) 
to 05$^{h}$40$^{m}$12$^{s}$ (85$\degr$.05). The lowest gray-scale level corresponds to $15\sigma$ 
 (where $1\sigma$ = 0.005629 K degree), with successive levels increasing in steps of $15\sigma$. 
In all the panels, a cyan dashed line shows the location of the ionizing star.}
\label{fig9}
\end{figure*}

\begin{figure*}
\epsscale{1.0}
\plotone{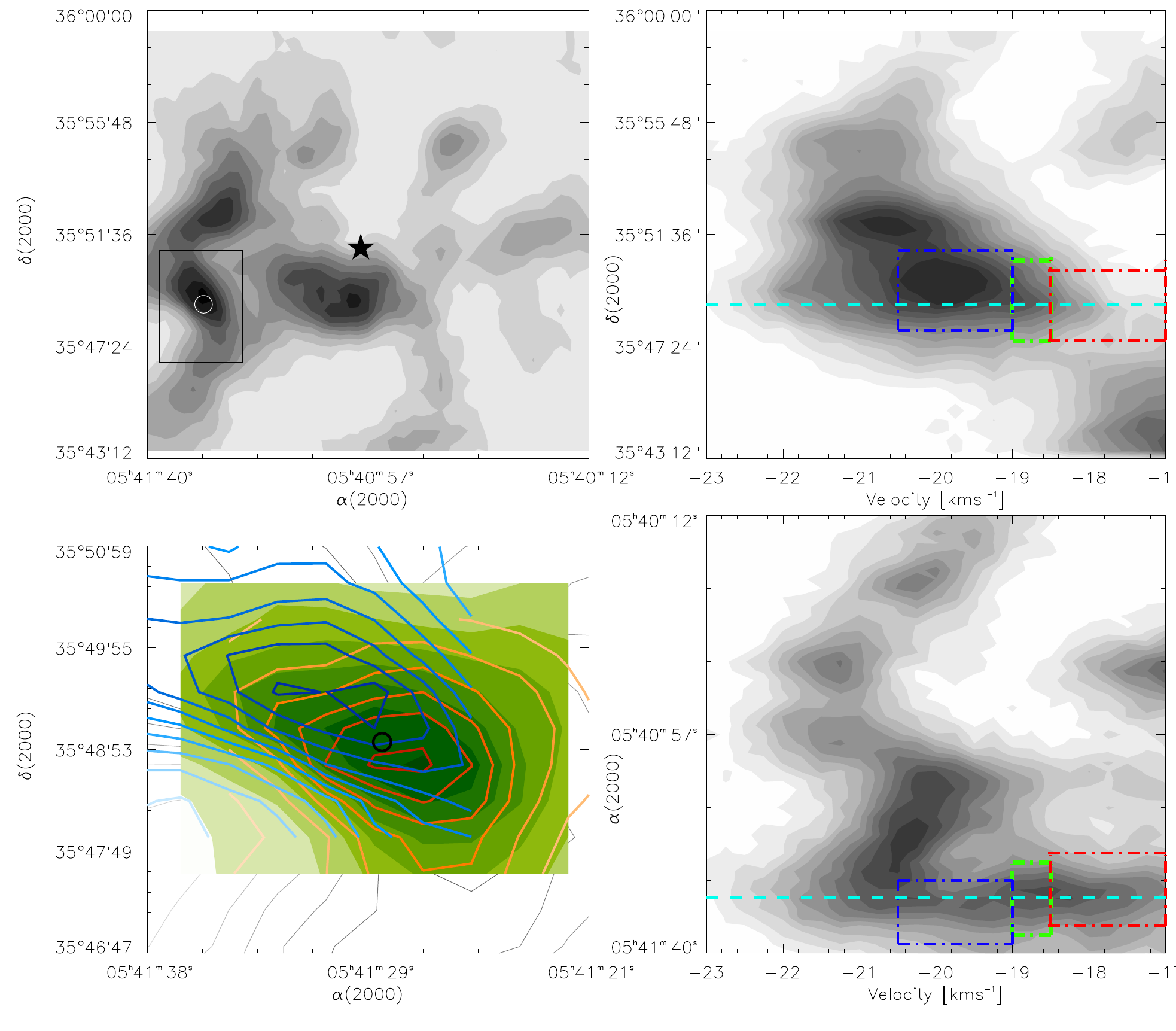}
\caption{\scriptsize Outflow signature in East~1 subregion. {\bf Top Left:} 
A contour map of integrated $^{13}$CO emission in the velocity range of $-$23 to $-$17 km s$^{-1}$.  
The lowest gray-scale level represents $20\sigma$, with successive levels 
increasing in steps of $20\sigma$. The channel RMS is 0.159 K km s$^{-1}$. 
The solid box in East~1 subregion is shown as a zoomed-in view in bottom left panel. 
The location of an O9.5V star is shown by a filled star. 
{\bf Bottom Left:} The selected area of contour map (in black color) is highlighted by a box in top left panel. 
Different color contours show the receding gas ($-$18.5  to  $-$17.0~km\,s$^{-1}$; red) 
and approaching gas ($-$20.5  to  $-$19.0~km\,s$^{-1}$; blue) with respect to that at 
rest ($-$19.0  to  $-$18.5~km\,s$^{-1}$; green). The spatial boundaries of these doppler components are shown in both the right panels. 
{\bf Top Right:} Declination-velocity map. 
{\bf Bottom Right:} Right Ascension-velocity map. 
In both the right panels, the contours are similar to those shown in Figure~\ref{fig9}.
The slices of receding, approaching, and rest gas are shown in red, blue, and green dashed boxes in both the right panels, respectively. 
The position of the probable driving source of outflow ($\alpha_{2000}$ = 05$^{h}$ 41$^{m}$ 29.5$^{s}$, 
$\delta_{2000}$ = +35$\degr$ 48$\arcmin$ 58$\farcs$7) is shown by an open circle in both the left panels. 
The cyan dashed line represents the position of the probable driving source of outflow in both the right panels.} 
\label{fig10} 
\end{figure*}

\begin{figure*}
\epsscale{1.0}
\plotone{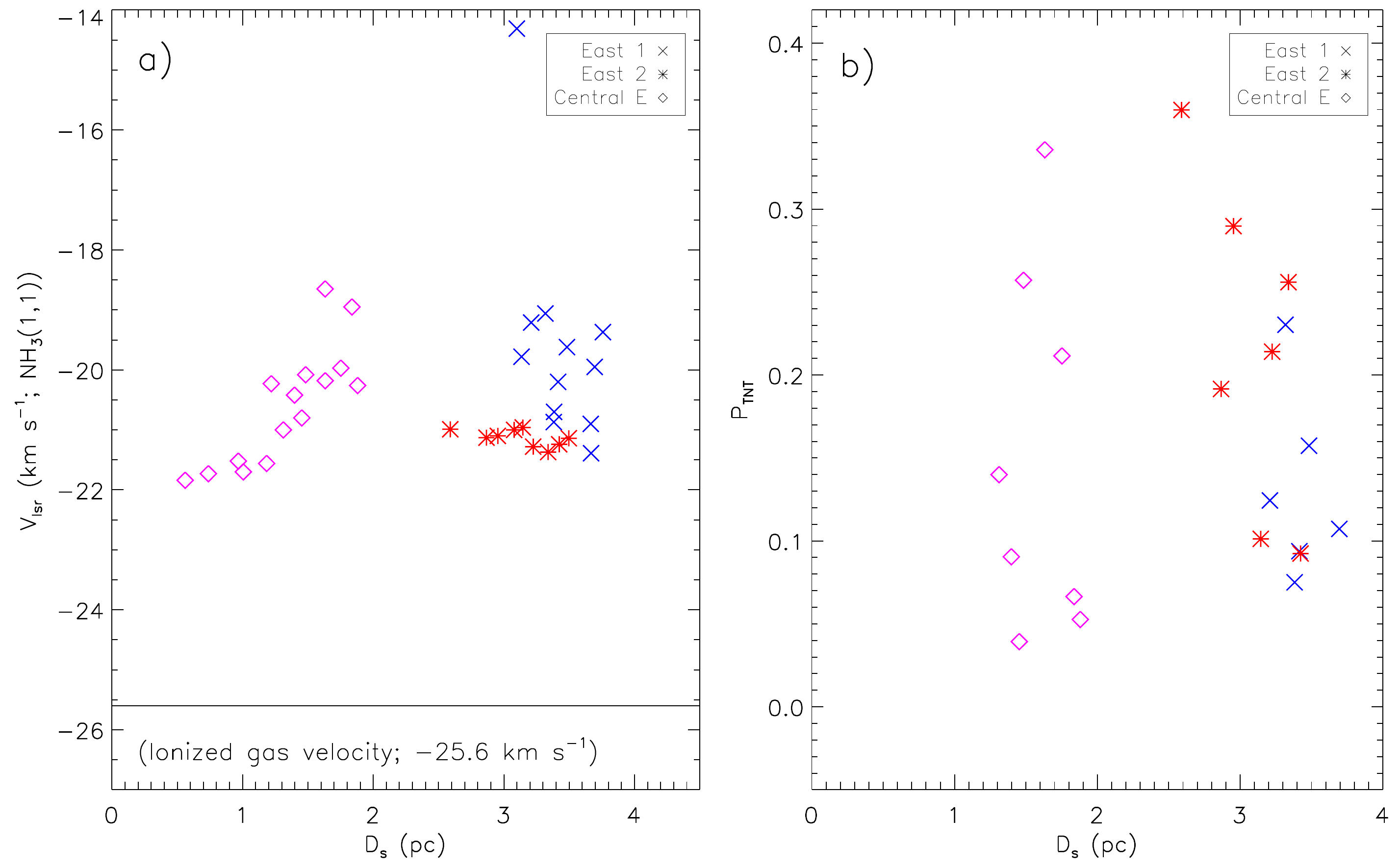}
\caption{\scriptsize a) Figure shows a variation of NH$_{3}$(1,1) radial velocity (V$_{lsr}$) in each subregion 
as a function of distance from the location of an O9.5V star (D$_{s}$) (see Figure~\ref{fig2a}b and Table~\ref{tab3}). 
b) Ratio of thermal to non-thermal gas pressure (P$_{TNT}$ = ${a^2_{s}}/{\sigma^2_{NT}}$) vs. D$_{s}$ (see Table~\ref{tab3}). 
The low value of P$_{TNT}$ corresponds to the dominant term in the denominator (i.e. non-thermal velocity dispersion ($\sigma_{NT}$)).} 
\label{fig15} 
\end{figure*}

\begin{figure*}
\epsscale{0.9}
\plotone{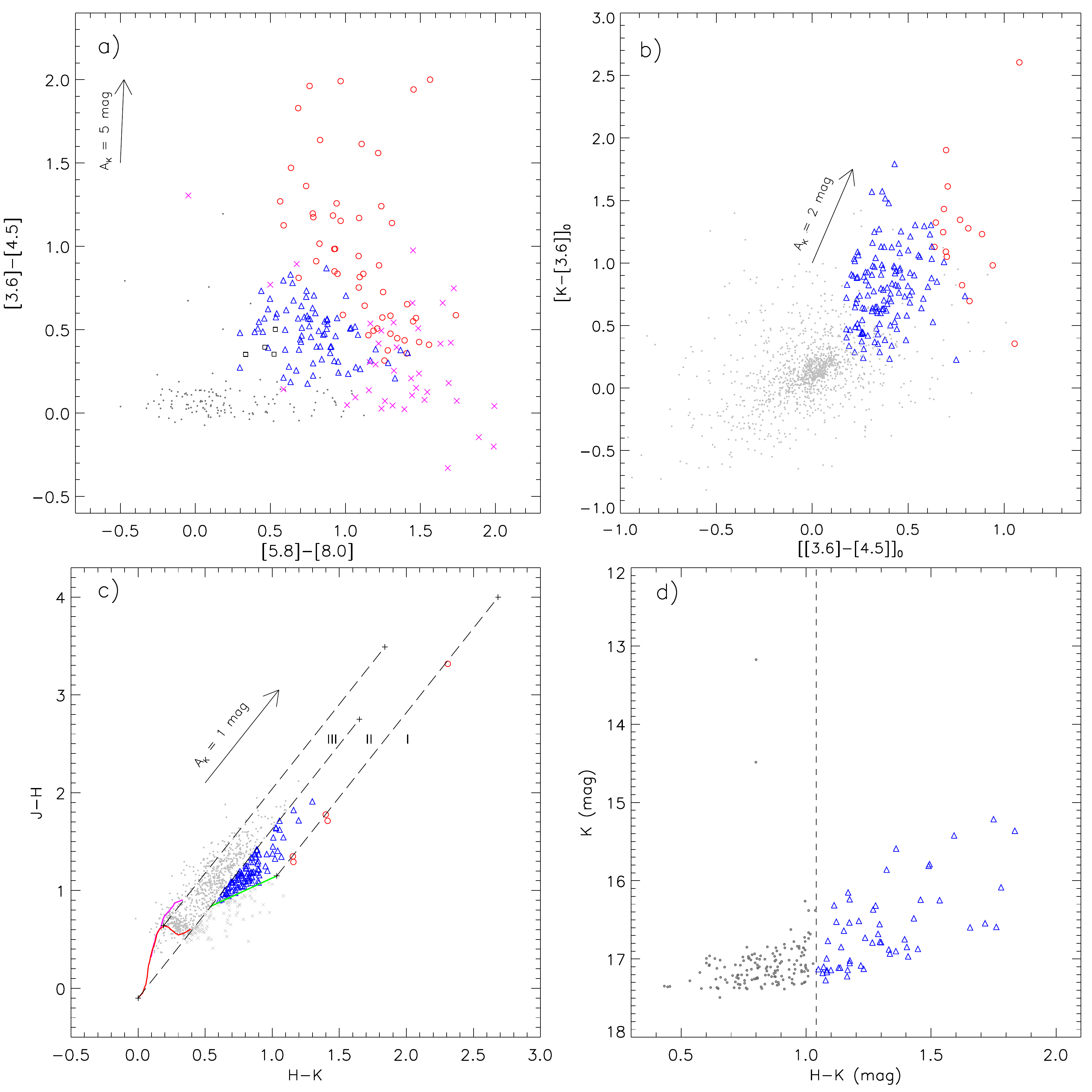}
\caption{\scriptsize a) {\it Spitzer}-IRAC color-color diagram ([3.6]$-$[4.5] vs. [5.8]$-$[8.0]) \citep[from][]{dewangan11}. 
The ``$\Box$'' and ``$\times$'' symbols indicate the Class~III sources and identified PAH-emission-contaminated apertures, respectively; 
b) The dereddened [K$-$[3.6]]$_{0}$ $vs$ [[3.6]$-$[4.5]]$_{0}$ color-color diagram using GPS and IRAC data; 
c) NIR color-color diagram (H$-$K vs J$-$H). The solid curves show the unreddened 
locus of main sequence stars (red) and giants (magenta) \citep[from][]{bessell88}. 
Classical T Tauri (CTTS) locus (in California Institute of Technology (CIT) system) \citep{meyer97} is 
drawn with a solid green line. The three parallel long-dashed lines represent 
the reddening vectors (with A$_{K}$ = 3 mag) of giants, main-sequence stars, and CTTS stars. 
The extinction vector A$_{K}$ = 1 mag is also shown in the diagram. 
The extinction vectors are drawn from \citet{indebetouw05} extinction laws. 
The color-color diagram is divided into three different subregions, namely ``I'', ``II'', and ``III''. 
The loci of the unreddened dwarf (Bessell \& Brett (BB) system), 
giant (BB-system) and CTTS (CIT system) are transfered into the 2MASS system using transformation 
equations given in \citet{carpenter01}; 
d) Color-magnitude diagram (H$-$K/K) of the sources detected in H and K bands. 
In all the panels, we show Class~I (red circles) and Class~II (open blue triangles) YSOs. 
In the first two panels, the arrow represents the extinction vector corresponding to the average extinction laws from \citet{flaherty07}. 
The dots (in gray color) show the stars with only photospheric emissions (see the text for YSOs selection conditions).}
\label{fig16}
\end{figure*}

\begin{figure*}
\epsscale{1.0}
\plotone{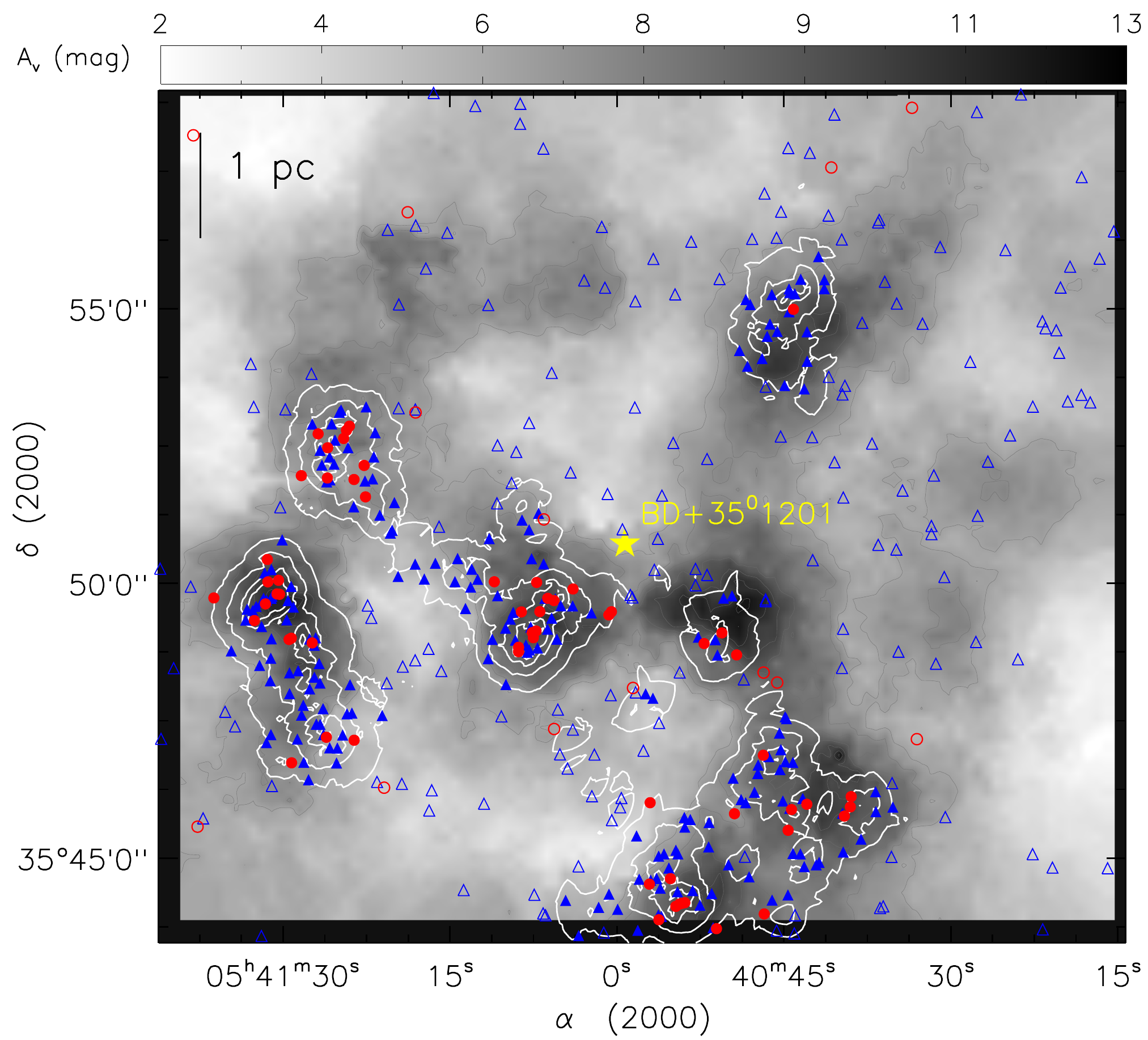}
\caption{\scriptsize Overlay of YSO surface density contours on the extinction map. 
The background map is similar to the one shown in Figure~\ref{fig2a}a. 
The surface density contours (in white color) are shown at 10, 20, 40, and 70 YSOs pc$^{-2}$, increasing from 
the outer to the inner regions. 
The positions of Class~I and Class~II YSOs identified within our selected region are shown by red circles and blue triangles, respectively. 
All the cluster members (YSOs having d$_{c}$ $\leq$ 0.453 pc) are shown by filled symbols (see text for details). 
The YSOs greater than d$_{c}$ ($>$ 0.453 pc) are marked by open symbols. 
The location of an O9.5V star is shown by a filled star. }
\label{fig17}
\end{figure*}

\begin{figure*}
\epsscale{1.0}
\plotone{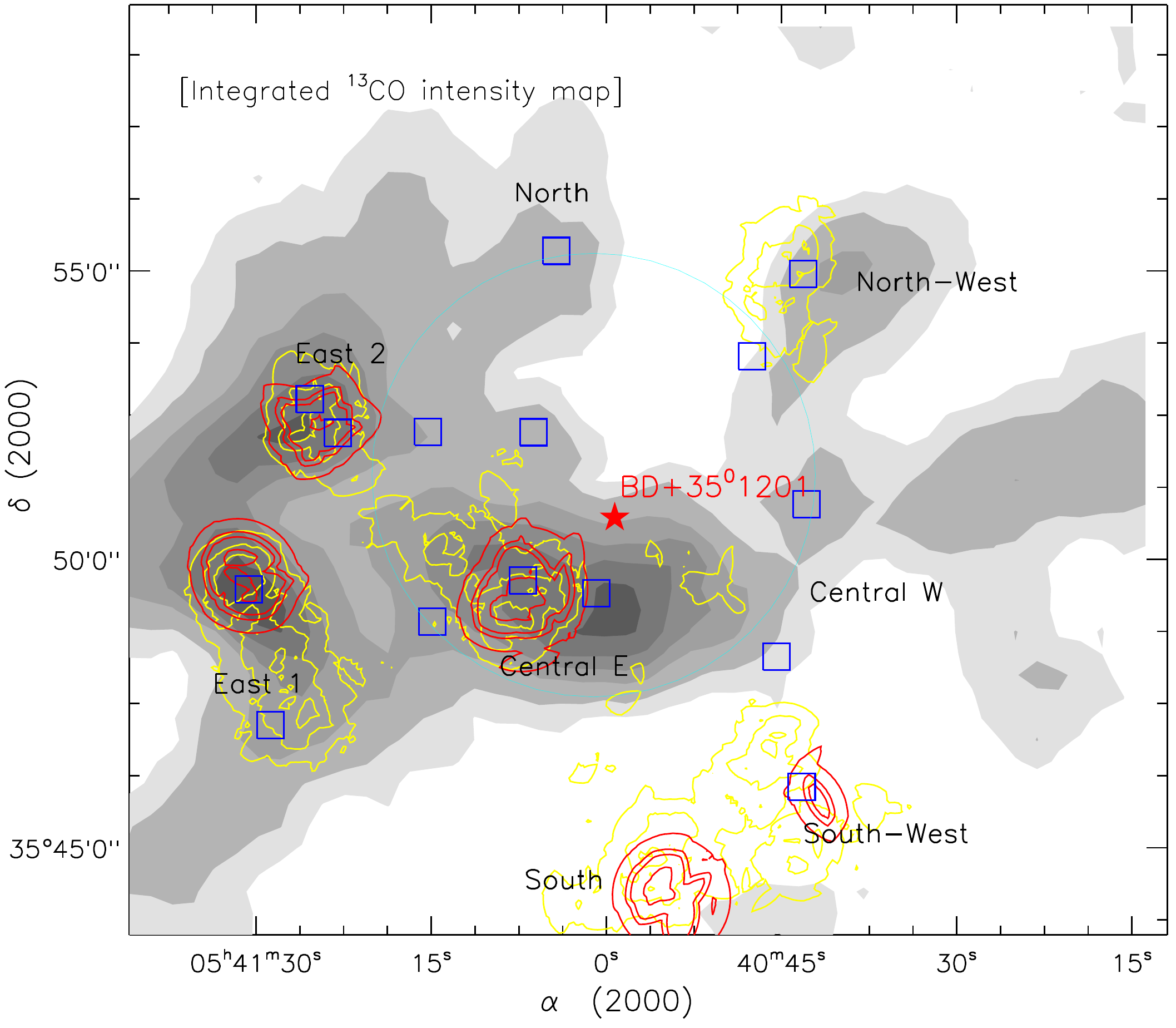}
\caption{\scriptsize Overlay of Class I and Class II YSO surface density contours on the integrated $^{13}$CO map. 
The background map is similar to the one shown in Figure~\ref{fig8}b. 
The surface density contours of Class I YSOs are drawn in red color with 5, 8, 10, and 20 YSO pc$^{-2}$. 
The Class II YSO surface density contours are overlaid in yellow color with 10, 20, and 40 YSO pc$^{-2}$. 
The location of an O9.5V star is shown by a filled star. 
The positions of dust clumps are shown by blue squares in the map (see Table~\ref{tab1}).} 
\label{fig18} 
\end{figure*}

\begin{figure*}
\epsscale{1.0}
\plotone{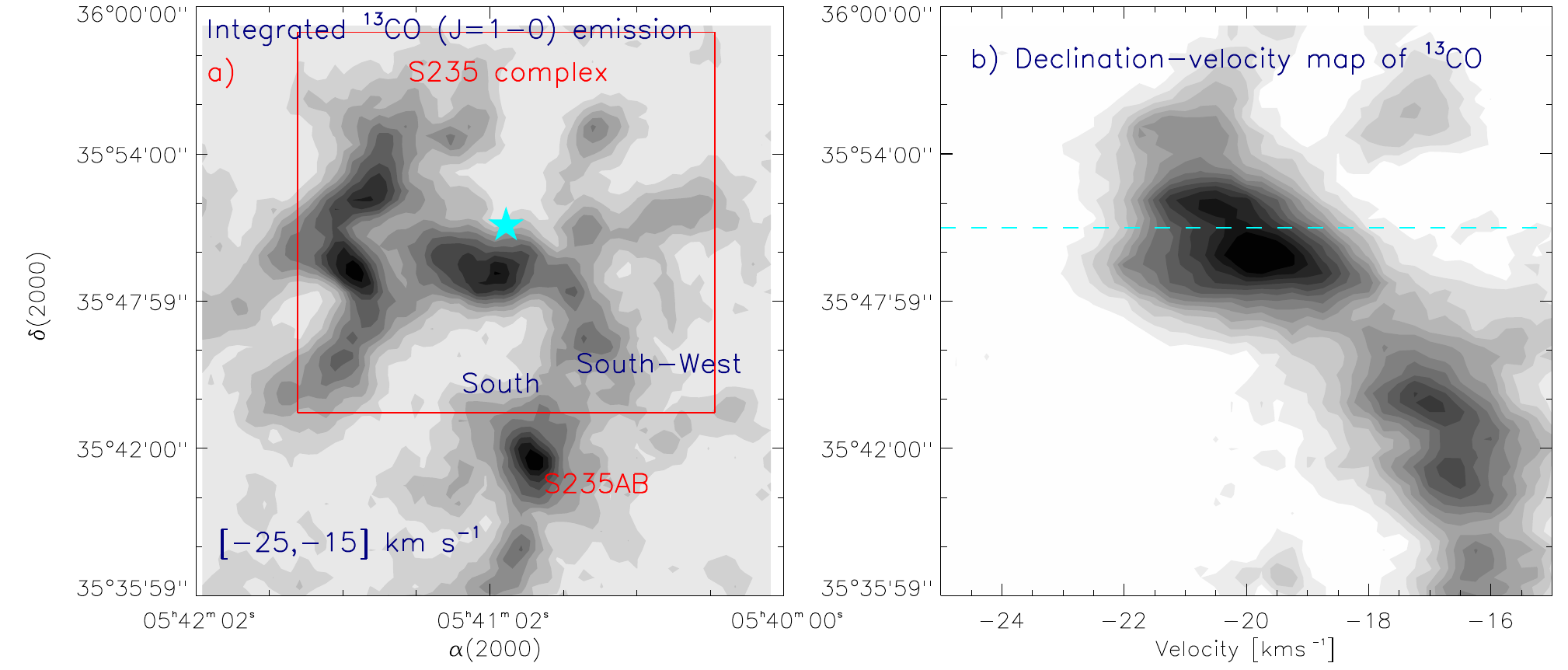}
\epsscale{0.65}
\plotone{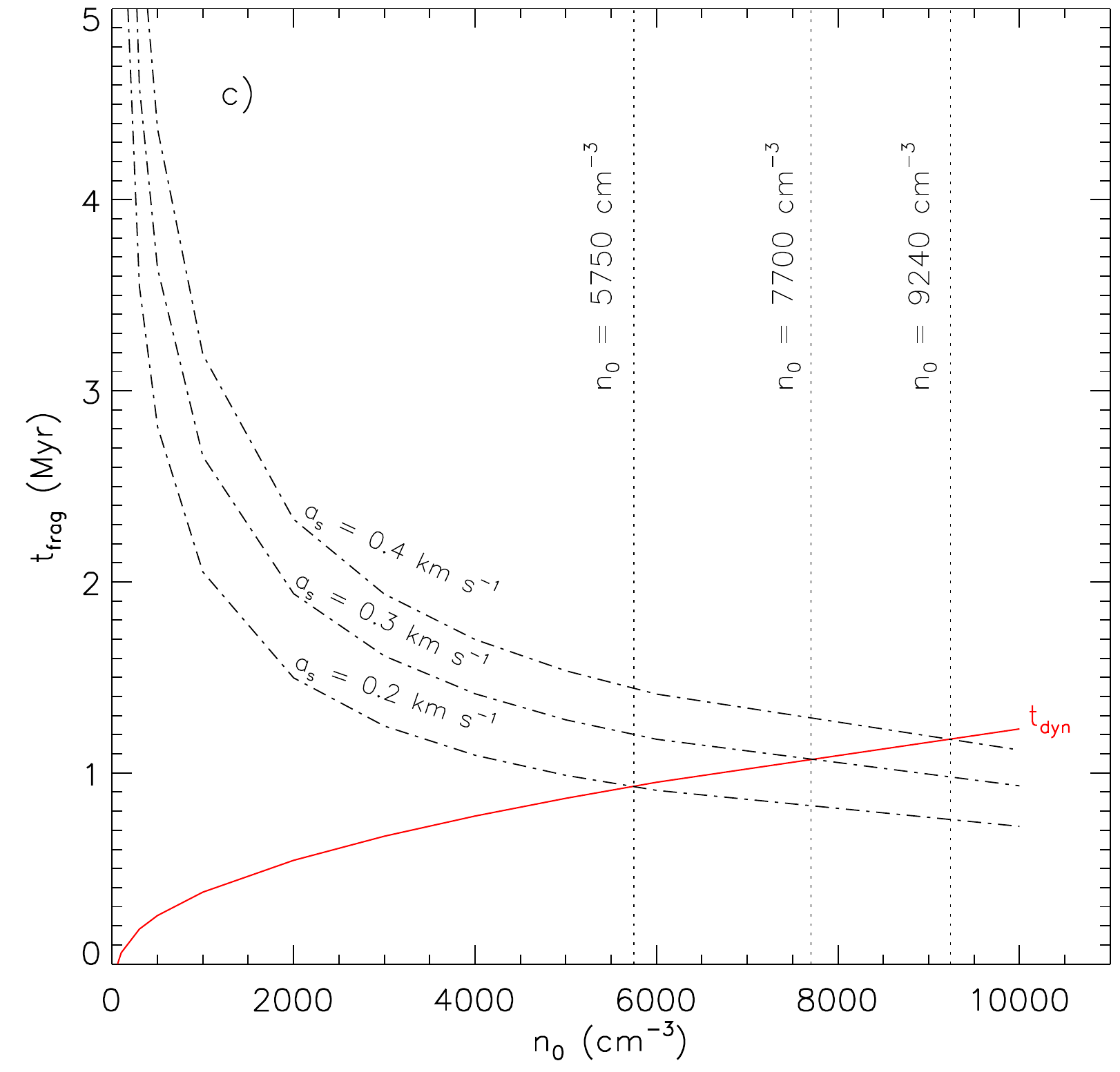}
\caption{\scriptsize a) The distribution of molecular gas toward the S235 complex and the S235AB region. 
A contour map of integrated $^{13}$CO emission in the velocity range of $-$25 to $-$15 km s$^{-1}$.  
The location of an O9.5V star is shown by a filled star. The solid box shows the field of the S235 complex, as shown in Figure~\ref{fig1a}a. 
The S235AB region is also marked in the map. b) Declination-velocity map. 
The position-velocity diagram suggests the possibility of interaction between the S235 molecular cloud and the S235AB molecular cloud.
c) Plot of fragmentation timescale (t$_{frag}$) and dynamical time (t$_{dyn}$) as function of initial density (n$_{0}$) of the ambient neutral medium.
The fragmentation timescale is computed for different sound speed of neutral gas ($a_{s}$) = 0.2, 0.3, and 0.4 km s$^{-1}$ (see Table~\ref{tab3}). 
Dotted lines depict the value of ``n$_{0}$" at which ``t$_{dyn}$" is equal to ``t$_{frag}$" for different ``$a_{s}$" values (see the text for more details).} 
\label{fig19a} 
\end{figure*}

Concerning the molecular gas distribution toward the South and South-West subregions (see Figure~\ref{fig7}), 
we suggest that these two subregions seem to be located at the interface between the S235 
molecular cloud ($-$23 km~s$^{-1}$ $< V_{lsr} < -$18 km~s$^{-1}$) 
and the S235AB molecular cloud ($-$17 km~s$^{-1}$ $< V_{lsr} < -$15 km~s$^{-1}$). 
As previously also reported by \citet{evans81a}, the extended star-forming region S235 is comprised of two velocity components 
at $-$20 and $-$17 km s$^{-1}$. The velocity information suggests that the molecular gas associated with S235AB region is red-shifted with respect to 
the S235 molecular cloud. In the integrated $^{12}$CO map, some molecular gas is seen toward the South and South-West subregions (see Figure~\ref{fig8}a), 
however the molecular gas is absent there in the integrated $^{13}$CO map (see Figure~\ref{fig8}b). 
Additionally, A$_{V}$ value is found significant there (see Figure~\ref{fig2a}a). 
Furthermore, the molecular gas is noticeably traced toward these subregions in 
a velocity range $-$17 km~s$^{-1}$ to $-$15 km~s$^{-1}$ (see Figure~\ref{fig7}). 
In Figures~\ref{fig19a}a and~\ref{fig19a}b, we present the distribution of molecular gas toward the 
``S235 complex" and the ``S235AB region". 
The CO map is integrated in the [-25,-15] km s$^{-1}$ velocity range (see Figure~\ref{fig19a}a). 
In Figure~\ref{fig19a}b, the p-v plot of $^{13}$CO gas reveals an almost broad bridge feature. 
The p-v plot shows that two peaks (a red shifted and a blue shifted; as mentioned above) are separated by 
lower intensity intermediated velocity emission.
The presence of such broad bridge feature in the p-v plot might indicate the signature of a collision between two clouds 
\citep[e.g.][]{haworth15a,haworth15b}. 
Therefore, there is a possibility of the formation of YSO clusters associated with 
South and South-West subregions by the interaction between these two clouds. 
In the present work, our analysis is mainly focused on the S235 complex, therefore the results related to 
the S235AB region are not presented in this work. 
Hence, a detailed investigation of the interaction between these two clouds is beyond the scope of present work. 
It is important to note that the broad bridge feature is not seen in the p-v maps of ``S235 complex" 
(see Figure~\ref{fig9}), suggesting the applicability of cloud-cloud collision scenario is unlikely in the S235 complex. 
Hence, we rule out the cloud-cloud collision process within the S235 complex.
\subsection{Triggered star formation}
\label{subsec:signp1}
In Section~\ref{subsec:feed}, we showed that the total pressure driven by a massive star is in 
equilibrium with the pressure exerted by the self-gravitating molecular cloud. 
In Section~\ref{subsec:phyp}, we discussed that the signature of the expanding H\,{\sc ii} region is evident. 
Additionally, we inferred that the densest subregions (i.e. East~1, East~2, and Central~E)  traced by 
NH$_{3}$ data are supersonic (see Section~\ref{subsec:amonia}). We also pointed out that the observed higher velocity dispersions  
could be due to the presence of outflows in the subregions. 
\citet{morgan10} studied 44 bright-rimmed clouds (BRCs) using NH$_{3}$ data and 
found that the potentially triggered samples of BRCs are associated with higher velocity
dispersions compared to non-triggered sources.  
In their samples, non-triggered sources were largely subsonic, whereas the triggered samples of BRCs were supersonic. 
These results indicate that the S235 complex could be a site of triggered star formation. 
This interpretation is further supported by the observed configuration as traced in the extinction map, 
dust continuum emission, molecular gas, ionized emission, and distribution of YSOs. 
Therefore, the on-going star formation in the S235 complex may have been influenced by the expanding H\,{\sc ii} region. 

Theoretically, there are two main scenarios discussed in the literature which explain the triggered star formation 
by the expansion of the H\,{\sc ii} region \citep{elmegreen98,deharveng05}: ``collect and collapse" 
\citep[see][]{elmegreen77,whitworth94,dale07} and radiation-driven implosion \citep[RDI; see][]{bertoldi89,lefloch94}. 
In the ``collect and collapse" scenario, massive and dense shell of cool neutral material can be accumulated 
around an expanding H\,{\sc ii} region, and star formation occurs when this material becomes gravitationally unstable. 
In the RDI model, the expanding H\,{\sc ii} region causes the instability and aids in the collapse of 
a pre-existing dense clump in the molecular cloud. 
\citet{kirsanova14} utilized NH$_{3}$ line data and suggested that the subregions (i.e. S235 East~2 and S235 Central) were formed via triggering by a ``collect-and-collapse" process. Furthermore, they suggested that S235 East~1 region was formed as a result of an interaction of the shock front from S235 complex  with a pre-existing dense clump.

In the present work, Bolocam data analysis revealed that the clump masses grow as one goes away from the location of the ionizing star. 
This pattern is observed only for those clumps which are located near the edges of the sphere-like 
shell (e.g. Central~E, East~2, North, and North-West subregions). 
It seems that the material has been collected on the edges of the sphere-like shell by the expanding H\,{\sc ii} region (also see Figure~\ref{fig2a}a), 
suggesting the ``collect and collapse" scenario applicable in the complex. 
In order to check the ``collect and collapse" process as triggering mechanism, 
we calculated the dynamical age (t$_{dyn}$) of the H\,{\sc ii} region and the fragmentation time scale (t$_{frag}$) 
following the equations given in \citet{dyson80} and \citet{whitworth94}, respectively. 
One can find more details of similar analysis in \citet{dewangan12}. 
The condition for the ``collect and collapse" process is t$_{dyn}$ $\geq$ t$_{frag}$. 
To find the values of ambient density (n$_{0}$=) for t$_{dyn}$ $\geq$ t$_{frag}$ condition, 
the diagram of t$_{frag}$ and t$_{dyn}$ as a function of ``n$_{0}$" is shown in Figure~\ref{fig19a}c.
The t$_{frag}$ is estimated for different ``a$_{s}$" values of 0.2, 0.3, and 0.4 km s$^{-1}$ (see Section~\ref{subsec:amonia} for ``a$_{s}$" estimation). 
Here, we use N$_{uv}$ = 4.9 $\times$ 10$^{47}$ s$^{-1}$ (see Section~\ref{subsec:feed}) for the 
entire radio emission of the region (spatial extent $\sim$6$\farcm$40 or 3.35 pc) 
and $\alpha_{B}$ = 2.6 $\times$ 10$^{-13}$ cm$^{3}$ s$^{-1}$ at T$_{e}$ = 10000 K. 
If the condition ``t$_{dyn}$ $\geq$ t$_{frag}$" is satisfied then the values of ``n$_{0}$" should be greater than 5750, 7700, and 9240 cm$^{-3}$ 
for different ``a$_{s}$" values of 0.2, 0.3, and 0.4 km s$^{-1}$, respectively (see Figure~\ref{fig19a}c). 
On the other hand, if n$_{0}$ $<$ 5750 cm$^{-3}$, then the condition ``t$_{dyn}$ $\geq$ t$_{frag}$" is not fulfilled. 
Consequently, the fragmentation of the molecular materials into clumps will not occur due to the ``collect and collapse" process in the complex.

Using $^{13}$CO, NH$_{3}$, H$_{2}$CO, HCO$^{+}$, and HCN line data, 
a significant density variation (i.e. 1 $\times$ 10$^{3}$ -- 2 $\times$ 10$^{4}$ cm$^{-3}$) has been reported 
in the S235 complex \citep{evans81a,kirsanova14}. \citet{nordh84} noted that the density structure of the S235 cloud is non-homogeneous with 
high density clumps engulfed in a medium with lower density. 
\citet{kirsanova14} adopted the value of volume density of 7000 cm$^{-3}$ to compute the age of the S235 H\,{\sc ii} region. 
In order to compare the previous results, following \citet{kirsanova14}, we also adopt the same value of volume density (i.e. n$_{0}$ = 7000 cm$^{-3}$) in the present work. 
Substituting N$_{uv}$, n$_{0}$, and an average radius of the H\,{\sc ii} region R $\approx$ 1.67 pc, 
we find t$_{dyn}$ $\sim$1 Myr. Our estimated t$_{dyn}$ value is in agreement with the work of \citet{kirsanova14}. 
The t$_{frag}$ is computed to be about 0.85 Myr, 1.1 Myr, and 1.3 Myr, respectively, 
for a$_{s}$ = 0.2, 0.3, and 0.4 km s$^{-1}$. 
We find that the dynamical age of the H\,{\sc ii} region (t$_{dyn} \approx $ 1 Myr) is comparable with the fragmentation 
time scale (t$_{frag} \approx $ 0.85--1.3 Myr) for n$_{0}$ = 7000 cm$^{-3}$ and a$_{s}$ = 0.2--0.4 km s$^{-1}$. 
In general, the average ages of Class~I and Class~II YSOs are $\sim$0.44 Myr and $\sim$1--3 Myr \citep{evans09}, respectively.  
Considering these ages, one can notice that the dynamical age of the H\,{\sc ii} region is comparable to the ages of YSOs. 
These calculations suggest that the molecular materials have been fragmented into several condensations around the sphere-like shell 
(also see Figure~\ref{fig17}). 
Therefore, the ``collect and collapse" scenario seems to be applicable in the S235 complex, which could be responsible for the origin of Central~E, 
East~2, North, North-West, and Central~W subregions. 

Note that East~1 subregion is far away from the sphere-like shell and is the most massive clump in the complex. 
Additionally, this subregion is separated by the eastern emission wall from the sphere-like shell and is in pressure equilibrium 
and is associated with Class~I and Class~II YSO clusters. 
This subregion is considered as the youngest star-forming site in the complex \citep[see][]{kirsanova14} and is supersonic. 
Therefore, it seems that the YSO clusters in East~1 subregion could be originated by 
compression of the pre-existing dense material by the expanding H\,{\sc ii} region through the RDI process. 
\section{Summary and Conclusions}
\label{sec:conc}
In this paper, we have made an extensive investigation of the S235 complex using the multi-wavelength data covering radio through NIR wavelengths. 
The results obtained in this work are based on the study of ionized emission, molecular emission, cold dust emission, 
embedded young populations, and various physical calculations. 
The important findings of this work are as follows:\\
$\bullet$ The most prominent structure in the S235 complex is the sphere-like shell morphology, 
as traced at wavelengths longer than 2 $\mu$m.\\ 
$\bullet$ The distribution of ionized emission observed in the GMRT 610 MHz and NVSS 1.4 GHz continuum maps is 
almost spherical. 
The ionizing photon flux values estimated at both the frequencies correspond to a single ionizing star of O9.5V spectral type. 
The location of the counterpart of this ionizing star (i.e. BD +35$^{o}$1201) appears at the 
peak of radio continuum emissions as well as approximately at the center of the sphere-like shell.\\
$\bullet$ The NIR extinction map depicts eight subregions (East~1, East~2, North, North-West, Central~W, Central~E, 
South-West, and South; having A$_{V}$ $>$ 8 mag), and five of them seem to be located in an almost 
regularly spaced manner along the sphere-like shell surrounding the ionized emission.\\ 
$\bullet$ Bolocam dust continuum emission at 1.1 mm is found toward all the eight subregions 
and provides more insight of the dense and cold regions of the S235 complex, which is lacking 
in the published sub-mm maps in the literature.\\ 
$\bullet$ Bolocam clump masses increase as one moves away from the location of the ionizing star.
This characteristic is found only for those clumps which are located near the edges of the shell.\\ 
$\bullet$ The position-velocity analysis of $^{12}$CO and $^{13}$CO emissions depicts an almost 
semi-ring like structure, indicating the signature of an expanding H\,{\sc ii} region.\\ 
$\bullet$ The pressure calculations (P$_{HII}$, $P_{rad}$, and P$_{wind}$) indicate that the photoionized gas associated with the S235 H\,{\sc ii} 
region can be considered as the major contributor for the feedback mechanism in the S235 complex.\\ 
$\bullet$ A cavity of empty molecular gas is investigated between North and North-West subregions, 
illustrating the interaction between the ionized gas  and the molecular gas.\\
$\bullet$ The mean values of Mach numbers inferred using NH$_{3}$ line data 
for three out of eight subregions (East~1, East~2, and Central~E) are 2.9, 2.3, and 2.9, suggesting 
these subregions are supersonic.\\ 
$\bullet$ The molecular outflows are evident in East~1, East~2, and Central~E subregions.\\ 
$\bullet$ The analysis of UKIDSS-NIR and {\it Spitzer}-IRAC photometry reveals 
a total of 435 YSOs, $\sim$59\% of which are found in clusters associated with the molecular cloud. \\
$\bullet$ In six subregions (including five located near the edges of the shell), the YSOs clusters are very well 
spatially correlated with the extinction, dust continuum emission, and molecular gas surrounding the ionized emission.\\
$\bullet$ The observed configuration as traced in the extinction map, dust continuum emission, molecular gas, 
ionized emission, and distribution of YSOs favours the triggered star formation in the S235 complex.\\  
$\bullet$ The dynamical age of the S235 H\,{\sc ii} region (t$_{dyn} \approx $ 1 Myr) is comparable with the fragmentation 
time scale (t$_{frag} \approx $ 0.85--1.3 Myr) of accumulated gas layers for ambient density (n$_{0}$) = 7000 cm$^{-3}$ and 
sound speed (a$_{s}$) = 0.2--0.4 km s$^{-1}$. \\

Considering all the observational evidences presented in this work, 
we conclude that the S235 complex can be considered as a promising nearby site of triggered star formation 
where both ``collect and collapse" and RDI processes seem to be applicable. 
Additionally, the star formation activity at the interface between the 
S235 molecular cloud and the S235AB molecular cloud due to the cloud-cloud collision appears likely.

\acknowledgments
We thank the anonymous referee for providing useful comments, which greatly improved the scientific contents of the paper. 
LKD acknowledges the financial support provided by the FCT (Portugal) grant SFRH/BPD/79741/2011 and 
the CONACYT(M\'{e}xico) grant CB-2010-01-155142-G3. AL acknowledges the CONACYT(M\'{e}xico) grant CB-2012-01-1828-41.
This work is based on data obtained as part of the UKIRT Infrared Deep Sky Survey. 
This publication made use of data products from the Two Micron All Sky Survey (a joint project of the University of Massachusetts and 
the Infrared Processing and Analysis Center / California Institute of Technology, funded by NASA and NSF), archival 
data obtained with the {\it Spitzer} Space Telescope (operated by the Jet Propulsion Laboratory, California Institute 
of Technology under a contract with NASA). 
The Canadian Galactic Plane Survey (CGPS) is a Canadian project with international partners. 
The Dominion Radio Astrophysical Observatory is operated as a national facility by the 
National Research Council of Canada. The Five College Radio Astronomy Observatory 
CO Survey of the Outer Galaxy was supported by NSF grant AST 94-20159. The CGPS is 
supported by a grant from the Natural Sciences and Engineering Research Council of Canada. 
We thank F. Navarete for providing the narrow-band H$_{2}$ image through the survey of 
extended H$_{2}$ emission from massive YSOs. 
We thank M.~S.~N. Kumar for initial discussion on the near-infrared data. 

\clearpage
\begin{table*}
\scriptsize
\setlength{\tabcolsep}{0.05in}
\centering
\caption{Bolocam clumps at 1.1 mm (see Figure~\ref{fig2a}). 
Table contains BGPS source designations, positions, distance from an O9.5V star (D$_{s}$), deconvolved effective radius (R$_{eff}$), 
integrated flux density (S$_{\nu}$), clump mass (M$_{g}$), column density (N(H$_{2}$)), and self-gravitating pressure ($P_{clump}$ 
$\approx$ $\pi G (M_{g}/ \pi R_{eff}^2$)$^2$). 
Clump masses were estimated using the integrated fluxes for a dust temperature = 20 K at a distance of 1.8 kpc. 
The positions of clumps located near the sphere-like shell are highlighted by superscript ``$\dagger$'' (see Figure~\ref{fig3a} and also see text for more details).}
\label{tab1}
\begin{tabular}{lcccccccccccccc}
\hline  
  ID                             & BGPS                      &  RA              &   Dec                 & D$_{s}$ &  R$_{eff}$   & S$_{\nu}$  & M$_{g}$         &   N(H$_{2}$)  &   $P_{clump}$        &    Subregion                  \\   
                                  &  Name                      & [2000]          &  [2000]               & (pc)       &  $\arcsec$(pc)          &  [Jy]            &  (M$_{\odot}$)  &   (10$^{22}$ cm$^{-2}$)   &   (10$^{-10}$ dynes\, cm$^{-2}$) &  \\   
\hline 						   	      	   							      
1                                & G173.570+02.749     &   05:40:42.9   &      +35:50:57     &  1.75	&     50.53(0.44)   &  0.549	 &    23.10    & 1.10  & 0.13 &  Central~W \\										       
2$^{\dagger}$             & G173.515+02.785     &   05:40:43.2   &	+35:54:57     &  2.80	 &     80.20(0.70)   &  1.345	  &    56.59    &  2.69&  0.12 &North-West\\
3                                & G173.640+02.707     &   05:40:43.3   &      +35:46:04     &  2.98	&     57.77(0.50)   &  1.429	 &    60.12    & 2.86 &  0.54 &South-West\\
4                                & G173.613+02.733     &   05:40:45.4   &      +35:48:18     &  1.94	&     59.10(0.52)   &  1.947	 &    81.92    &3.90 &   0.85 &Central~W \\
5$^{\dagger}$             & G173.543+02.785     &   05:40:47.6   &	+35:53:31     &  1.92	 &      --      &  0.193    &     8.12     & 0.39 &     --  &North-West  \\
6$^{\dagger}$             & G173.625+02.787     &   05:41:00.9   &	+35:49:24     &  0.71	 &     95.09(0.83)   &  5.128	  &   215.75   & 10.30  & 0.91&Central~E \\
7$^{\dagger}$             & G173.546+02.849     &   05:41:04.3   &	+35:55:21     &  2.48	 &     37.49(0.33)   &  0.438	  &    18.43    & 0.88 &  0.26&North \\
8                                & G173.594+02.827     &   05:41:06.2   &	    +35:52:12     &  1.07    &      --      &  0.482	&    20.28    & 0.96   &   --  & --  \\
9$^{\dagger}$             & G173.633+02.807     &   05:41:07.2   &	+35:49:38     &  1.01	 &     93.12(0.81)   &  5.978	  &   251.53   & 12.0  & 1.36 &Central~E\\
10$^{\dagger}$           & G173.657+02.823     &   05:41:14.9   &      +35:48:55     &  1.91	&	--     &  0.255	   &    10.73    & 0.51  &    --  & Central~E\\
11                              & G173.610+02.853     &   05:41:15.3   &	   +35:52:13     &  1.87    &	    --     &  0.216    &     9.10     & 0.43  &    -- & --    \\
12$^{\dagger}$           & G173.625+02.875     &   05:41:23.0   &      +35:52:12     &  2.63	&	--     &  1.041	   &    43.80    & 2.08  &    --  &   East 2   \\
13$^{\dagger}$           & G173.621+02.887     &   05:41:25.4   &      +35:52:47     &  2.97	&      44.71(0.39)  &  2.087	 &    87.81    &  4.18 & 3.09  &  East 2   \\
14                              & G173.706+02.847     &   05:41:28.8   &      +35:47:08     &  3.65    &	--    &  0.166	  &     6.98     &  0.33 &    --  &  East 1   \\
15                              & G173.676+02.873     &   05:41:30.6   &      +35:49:29     &  3.39    &      80.34(0.70)  &  6.768	&   284.76   & 13.5  & 3.13&   East 1  \\
\hline          
\end{tabular}
\end{table*}
\begin{table*}
\scriptsize
\setlength{\tabcolsep}{0.05in}
\centering
\caption{\scriptsize Physical parameters derived from NH$_{3}$ line data \citep[from][]{kirsanova14}. 
Table contains the information about three subregions (e1 = East1, e2 = East2, and cn = Central~E), 
positions, distance from an O9.5V star (D$_{s}$), kinematical temperature (T$_{kin}$), volume density (n(H$_{2}$)), NH$_{3}$ (1,1) line 
width ($\Delta V$), NH$_{3}$ (1,1) line velocity (V$_{lsr}$), 
sound speed ($a_{s}$), thermal velocity dispersion ($\sigma_{T}$), non-thermal velocity dispersion ($\sigma_{NT}$), and 
ratio of thermal to non-thermal gas pressure (P$_{TNT}$ = ${a^2_{s}}/{\sigma^2_{NT}}$) (see text for more details and also 
Figures~\ref{fig2a}b and~\ref{fig15}).}
\label{tab3}
\begin{tabular}{lccccccccccccc}
\hline 										    	        			      
  Region     &   RA 	   &      Dec     &    D$_{s}$               &     T$_{kin}$   &    n(H$_{2}$)	   &	$\Delta V$ &   V$_{lsr}$          & $a_{s}$        & $\sigma_{T}$&$\sigma_{NT}$ & P$_{TNT}$  \\ 
          &   [2000]	   &    [2000]    &    (pc)                     &     (K)      &   (10$^{3}$ cm$^{-3}$)   &   (km s$^{-1}$) &   (km s$^{-1}$)   &   (km s$^{-1}$)  &   (km s$^{-1}$) &   (km s$^{-1}$) &  (${a^2_{s}}/{\sigma^2_{NT}}$) \\ 
\hline 
   e1     & 	5:41:33.8  &    +35:51:07    &	    3.67       	 &	--   &  	 --   &     2.14 &  -21.39	       &    --    &    --      &   --	    &	--		      \\ 
   e1	  & 	5:41:33.8  & 	+35:50:27    &      3.66       	 &	--   &       --   &	2.34 &  -20.90  	       &    --      &    --    &   --	    &	--		      \\ 
   e1  	  & 	5:41:33.8  & 	+35:49:47    &      3.70      	 &    21.20  &       --   &	1.95 &  -19.95  	       &  0.27  &  0.10      & 0.82	    & 0.11		      \\ 
   e1 	  & 	5:41:33.8  & 	+35:49:07    &      3.76       	 &	--   &       --   &	1.52 &  -19.37  	       &    --     &    --     &   --	    &	--		      \\ 
   e1 	  & 	5:41:31.1  & 	+35:51:07    &      3.38       	 &	--   &       --   &	1.06 &  -20.70  	       &    --    &    --     &   --	    &	--		      \\ 
   e1     & 	5:41:31.1  & 	+35:50:27    &      3.38      	 &    20.73  &      18    &	2.30 &  -20.87  	       &  0.27   &  0.10     & 0.97	    & 0.08		      \\ 
   e1  	  & 	5:41:31.1  & 	+35:49:47    &      3.41      	 &    18.57  &      19    &	1.95 &  -20.20  	       &  0.25  &  0.09      & 0.82	    & 0.09		      \\ 
   e1  	  & 	5:41:31.1  & 	+35:49:07    &      3.48      	 &    31.53  &      19    &	1.97 &  -19.62  	       &  0.33  &  0.12      & 0.83	    & 0.16		      \\ 
   e1  	  & 	5:41:28.4  & 	+35:50:27    &      3.10       	 &	--   &       --   &	2.69 &  -14.31  	       &    --    &    --     &   --	    &	--		      \\ 
   e1 	  & 	5:41:28.4  & 	+35:49:47    &      3.14       	 &	--   &       --   &	1.54 &  -19.78  	       &    --   &    --      &   --	    &	--		      \\ 
   e1 	  & 	5:41:28.4  & 	+35:49:07    &      3.21      	 &    19.31  &      13    &	1.73 &  -19.21  	       &  0.26   &  0.10     & 0.73	    & 0.12		      \\ 
   e1	  & 	5:41:28.4  & 	+35:48:27    &      3.32      	 &    18.99  &       --   &	1.27 &  -19.06  	       &  0.25   &  0.10     & 0.53	    & 0.23		      \\ 
   e2	  & 	5:41:31.1  & 	+35:52:27    &      3.50       	 &	--   &       --   &	1.68 &  -21.14  	       &    --      &    --   &   --	    &	--		      \\ 
   e2	  & 	5:41:31.1  & 	+35:51:47    &      3.42      	 &    17.56  &       --   &	1.91 &  -21.24  	       &  0.24   &  0.09     & 0.81	    & 0.09		      \\ 
   e2	  & 	5:41:28.4  & 	+35:51:47    &      3.14      	 &    16.69  &       --   &	1.78 &  -20.96  	       &  0.24   &  0.09     & 0.75	    & 0.10		      \\ 
   e2	  & 	5:41:28.4  & 	+35:52:27    &      3.22      	 &    17.69  &       --   &	1.27 &  -21.28  	       &  0.25    &  0.09    & 0.53	    & 0.21		      \\ 
   e2	  & 	5:41:28.4  & 	+35:53:07    &      3.34      	 &    28.56  &       3    &	1.48 &  -21.37  	       &  0.31   &  0.12     & 0.62	    & 0.26		      \\ 
   e2	  & 	5:41:25.8  & 	+35:51:47    &      2.87      	 &    17.68  &       --   &	1.34 &  -21.13  	       &  0.25   &  0.09     & 0.56	    & 0.19		      \\ 
   e2	  & 	5:41:25.8  & 	+35:52:27    &      2.95     	 &    20.81  &       --   &	1.19 &  -21.10  	       &  0.27   &  0.10     & 0.50	    & 0.29		      \\ 
   e2	  & 	5:41:25.8  & 	+35:53:07    &      3.08       	 &	--   &       --   &	1.47 &  -21.00  	       &    --      &    --   &   --	    &	--		      \\ 
   e2	  & 	5:41:23.1  & 	+35:51:47    &      2.59      	 &    36.45  &       --   &	1.42 &  -20.99  	       &  0.35   &  0.13     & 0.59	    & 0.36		      \\ 
   cn	  & 	5:41:15.1  & 	+35:49:47    &      1.75      	 &    22.48  &       --   &	1.44 &  -19.97  	       &  0.28  &  0.10      & 0.60	    & 0.21		      \\ 
   cn	  & 	5:41:15.1  & 	+35:49:07    &      1.88      	 &    22.25  &      1	  &	2.84 &  -20.26  	       &  0.28  &  0.10      & 1.20	    & 0.05		      \\ 
   cn	  & 	5:41:12.4  & 	+35:49:47    &      1.48      	 &    27.92  &      4	  &	1.46 &  -20.08  	       &  0.31   &  0.12     & 0.61	    & 0.26		      \\ 
   cn	  & 	5:41:12.4  & 	+35:49:07    &      1.63      	 &    21.61  &      6	  &	1.13 &  -20.18  	       &  0.27  &  0.10      & 0.47	    & 0.34		      \\ 
   cn	  & 	5:41:12.4  & 	+35:48:27    &      1.84      	 &    14.32  &       --   &	2.03 &  -18.95  	       &  0.22  &  0.08      & 0.86	    & 0.07		      \\ 
   cn	  & 	5:41:09.8  & 	+35:49:47    &      1.22       	 &     --    &       --   &	2.41 &  -20.23  	       &    --     &    --    &   --	    &	--		      \\ 
   cn	  & 	5:41:09.8  & 	+35:49:07    &      1.40      	 &    23.65  &       --   &	2.24 &  -20.42  	       &  0.28  &  0.11      & 0.95	    & 0.09		      \\ 
   cn	  & 	5:41:09.8  & 	+35:48:27    &      1.63      	 &     --    &       --   &	0.82 &  -18.65  	       &    --   &    --      &   --	    &	--		      \\ 
   cn	  & 	5:41:07.1  & 	+35:49:47    &      0.97       	 &     --    &       --   &	3.06 &  -21.52  	       &    --    &    --     &   --	    &	--		      \\ 
   cn	  & 	5:41:07.1  & 	+35:49:07    &      1.18       	 &     --    &       --   &	1.66 &  -21.56  	       &    --   &    --      &   --	    &	--		      \\ 
   cn	  & 	5:41:07.1  & 	+35:48:27    &      1.45      	 &   13.41   &      5	  &	2.55 &  -20.80  	       &  0.21  &  0.08       & 1.08	    & 0.04		      \\ 
   cn	  & 	5:41:04.4  & 	+35:49:47    &      0.74       	 &     --    &       --   &	1.54 &  -21.73  	       &    --       &    --  &   --	    &	--		      \\ 
   cn	  & 	5:41:04.4  & 	+35:49:07    &      1.01       	 &     --    &       --   &	1.87 &  -21.70  	       &    --       &    --  &   --	    &	--		      \\ 
   cn	  & 	5:41:04.4  & 	+35:48:27    &      1.31      	 &   15.44   &      2	  &	1.46 &  -21.00  	       &  0.23 &  0.09       & 0.61	    & 0.14		      \\ 
   cn	  & 	5:41:01.8  & 	+35:49:47    &      0.56      	 &   43.04   &      5	  &	1.07 &  -21.84  	       &  0.38 &  0.14       & 0.43	    & 0.79		      \\ 
\hline          		
\end{tabular}			
\end{table*}			
\begin{table*}
\scriptsize
\setlength{\tabcolsep}{0.01in}
\centering
\caption{Pressure components driven by an O9.5V star on different subregions of the S235 complex. 
The pressure exerted by the self-gravitating molecular cloud in each subregion is estimated 
to be $\approx$0.1--3.0 $\times$ 10$^{-10}$ dynes cm$^{-2}$ (see text for more details).}
\label{tab4}
\begin{tabular}{lccccccccccccc}
\hline 										    	        			      
 Pressure                      &   East 1 	           &    East 2                   &       Central~E                  &     North	                &   North-West    &    Central~W  &    South-West \\ 
   (dynes\, cm$^{-2}$)      &  at D$_{s}$ = 3.4 pc 	   &   at D$_{s}$ = 3.0 pc     &    at D$_{s}$ = 1.5 pc       &     at D$_{s}$ = 2.6 pc     &     at D$_{s}$ = 2.7 pc   &  at D$_{s}$ = 1.9 pc &  at D$_{s}$ = 3.0 pc \\ 
\hline 
 P$_{HII}$  & (7.8$\pm$0.2)$\times$10$^{-11}$  & (9.4$\pm$0.2)$\times$10$^{-11}$  &  (2.7$\pm$0.07)$\times$10$^{-10}$  &  (1.2$\pm$0.03)$\times$10$^{-10}$  &  (1.1$\pm$0.03)$\times$10$^{-10}$  &  (1.9$\pm$0.05)$\times$10$^{-10}$  &  (9.4$\pm$0.2)$\times$10$^{-11}$ \\ 
 P$_{rad}$    & (7.4$\pm$0.2)$\times$10$^{-12}$	  & (9.5$\pm$0.5)$\times$10$^{-12}$  &  (3.8$\pm$0.2)$\times$10$^{-11}$  &  (1.3$\pm$0.06)$\times$10$^{-11}$  &  (1.2$\pm$0.06)$\times$10$^{-11}$  &  (2.4$\pm$0.1)$\times$10$^{-11}$  &  (9.5$\pm$0.5)$\times$10$^{-12}$ \\ 
 P$_{wind}$  & (1.1$\pm$0.08)$\times$10$^{-14}$   & (1.4$\pm$0.1)$\times$10$^{-14}$  &  (5.6$\pm$0.4)$\times$10$^{-14}$  &  (1.9$\pm$0.1)$\times$10$^{-14}$  &  (1.7$\pm$0.1)$\times$10$^{-14}$  &  
 (3.5$\pm$0.2)$\times$10$^{-14}$  &  (1.4$\pm$0.1)$\times$10$^{-14}$ \\ 
 P$_{total}$  & (0.9$\pm$0.02)$\times$10$^{-10}$	  & (1.0$\pm$0.02)$\times$10$^{-10}$  &  (3.0$\pm$0.07)$\times$10$^{-10}$  &  (1.3$\pm$0.03)$\times$10$^{-10}$  &  (1.2$\pm$0.03)$\times$10$^{-10}$  &  (2.1$\pm$0.05)$\times$10$^{-10}$  &  (1.0$\pm$0.02)$\times$10$^{-10}$ \\ 
\hline          		
\end{tabular}			
\end{table*}			
%

\end{document}